\documentclass[12pt,preprint]{aastex} %% \documentclass[preprint2]{aastex}
\usepackage{graphicx}
\usepackage{amssymb}
\usepackage{amsmath}
\newcommand{\be}{\begin{equation}}
\newcommand{\ee}{\end{equation}}
\newcommand{\msun}{\mathrm{M_{\sun}}}

\shorttitle{Late-Time Origin of Extremely-Metal Poor Stars} 
\shortauthors{Trenti \& Shull}

\begin{document}

%% LaTeX will automatically break titles if they run longer than
%% one line. However, you may use \\ to force a line break if

\title{Extremely-Metal Poor Stars in the Milky Way: \\
  A Second Generation Formed after Reionization}

%% Use \author, \affil, and the \and command to format
%% author and affiliation information.
%% Note that \email has replaced the old \authoremail command
%% from AASTeX v4.0. You can use \email to mark an email address
%% anywhere in the paper, not just in the front matter.
%% As in the title, use \\ to force line breaks.

\author{Michele Trenti and J.~Michael Shull} 
\affil{University of Colorado, CASA, Dept.  of Astrophysical \& 
Planetary Sciences, 389-UCB, Boulder, CO 80309, USA} 
\email{trenti@colorado.edu, michael.shull@colorado.edu}

%% Notice that each of these authors has alternate affiliations, which
%% are identified by the \altaffilmark after each name.  Specify alternate
%% affiliation information with \altaffiltext, with one command per each
%% affiliation.

%-------------------------------------------------%
\begin{abstract}
  Cosmological simulations of Population~III star formation suggest an
  initial mass function (IMF) biased toward very massive stars ($M \gtrsim
  100~ \mathrm{M_{\sun}}$) formed in minihalos at redshift $z \gtrsim 20$, 
  when the cooling is driven by molecular hydrogen.  However, this result 
  conflicts with observations of extremely-metal poor (EMP) stars in the 
  Milky Way halo, whose r-process elemental abundances appear to be incompatible 
  with those expected from very massive Population~III progenitors. We propose 
  a new solution to the problem in which the IMF of second-generation stars 
  formed at $z \gtrsim 10$, before
  reionization, is deficient in sub-solar mass stars, owing to the high 
  cosmic microwave background temperature floor. The observed EMP 
  stars are formed preferentially at $z \lesssim 10$ in pockets of gas 
  enriched to metallicity $Z \gtrsim 10^{-3.5}~Z_{\sun}$ by winds from 
  Population~II stars.  Our cosmological simulations of dark matter halos
  like the Milky Way show that current samples of EMP stars can only 
  constrain the IMF of late-time Population~III stars, formed at 
  $z\lesssim 13$ in halos with virial temperature $T_{\rm vir} \sim 10^4$~K.  
  This suggests that pair instability supernovae were not produced primarily 
  by this population.  To begin probing the IMF of Population~III stars formed at 
  higher redshift will require a large survey, with at least $500$ and probably 
  several thousand EMP stars of metallicities $Z \approx 10^{-3.5}~Z_{\sun}$. 

\vspace{3cm}

\end{abstract}

\keywords{galaxies: high-redshift - Galaxy: evolution  - intergalactic
medium - stars: abundances - cosmology: theory - stars: formation }

%%%%%%%%%%%%%%%%%%%%%%%%%%%%%%%%%%%%
\section{Introduction}\label{sec:intro}

According to simulations, the first stars in the Universe began
forming 50--100 million years after the Big Bang (redshifts $z =
30-50$) in mini-halos with virial temperatures $T_{\rm vir} = 10^3$ K
\citep{tegmark97,yoshida03,naoz06,ts07a}. To fix the semantics of
early stellar populations, we define Population~II (stars and gas) as
material enriched to metallicities $Z > Z_{\rm crit} \approx 
10^{-3.5} Z_{\sun}$, at which point the star-formation mode and initial 
mass function (IMF) are altered through radiative cooling by metals, 
which dominate that of molecular hydrogen. The equality of the 
dynamical time and cooling time in a collapsing gas cloud marks the 
transition from a typical fragmentation mass of $\sim 10^3~M_{\sun}$ 
to a lower mass $M \lesssim 10^2~M_{\sun}$, whose exact value depends 
on possible coupling with the Cosmic Microwave Background (CMB) 
temperature (\citealt{smith_b09}; see also \citealt{brommloeb03}, 
\citealt{omukai05}). Therefore we define Population~III as pristine, 
high-redshift stars formed from gas at zero or near zero 
($Z < Z_{\rm crit}$) metallicity \citep{bromm01a,bromm04,santoro06,smith_b09}.

Initially very rare (one star per Gpc$^3$ at $z\approx 50$), these
early generations of stars form at rates that increase rapidly in number 
toward lower redshift until they are eventually suppressed by
radiative and chemical feedback (see \citealt{trenti09b} and
references therein). The numerical simulations further suggest that
the first-generation (Population~III) stars formed in isolation, with
an IMF with characteristic scale $\sim 100~ \msun$ consistent with
the fragmentation mass derived from simple cooling models (see
\citealt{abel02,bromm03,bromm04,yoshida06,oshea07,oshea08}). This
characteristic mass is set primarily by the physics of hydrogen
cooling, as these stars lack metals and dust. Molecular hydrogen
(H$_2$) can only cool the gas down to $\sim 200$ K \citep{galli98}, at
which point the gas reaches a density $\sim 10^4~ \mathrm{cm^{-3}}$
with corresponding Jeans mass $M_J \sim 10^3~ \msun$
\citep{yoshida06}. High accretion rates onto the protostellar core
($10^{-4}$ to $10^{-2}~ \msun~\mathrm{yr^{-1}}$, see
\citealt{oshea07}) suggest that their final mass is likely to be
$\gtrsim 100~\msun$. Radiative feedback will eventually shut down the
accretion \citep{tan04}, but this phase has not yet been simulated
self-consistently.

In the presence of a strong flux in the H$_2$ Lyman-Werner (LW) bands, 
with energies between 11.15--13.6 eV, the primary (H$_2$) coolant of 
minihalos is photo-dissociated \citep{lepp83,haiman97,machacek03}. Under 
these conditions, the minimum halo mass required for cooling increases from
$\sim 10^6~ \msun$ up to $\sim 10^8~ \msun$, reflecting a rise in
virial temperature from $T_{\rm vir}\sim 10^3$~K to $\sim 10^4$~K 
\citep{machacek03,greif06}. In these larger halos, small
clusters of Population III stars can form. The mass of the
gas at the halo center is typically larger than in the case of
a minihalo collapse, so a few tens of Jeans masses might be present.
Because these second-generation stars form relatively late, at 
$z \lesssim 15$, some of them can be made out of reionized gas with 
correspondingly smaller masses ($\sim 40~ \msun$, see \citealt{yoshida06}).

A direct detection of Population III stars will be extremely challenging, 
even with the next generation of telescopes and instruments.   A $100~
\msun$ Population III star has an observed magnitude $m_{AB} \gtrsim
38$ at $z>6$, whereas the {\it James Webb Space Telescope} (JWST) is 
limited to $m_{AB} \sim 31$.  Thus, JWST will only be able to observe 
such stars in the unlikely event they form in large clusters \citep{trenti09b}. 
Supernovae from massive Population III stars are bright enough to be 
seen at essentially any redshift, but their observed rate is low.
Therefore, a combination of deep multi-epoch imaging and large area 
coverage is needed \citep{weinmann05,trenti09b}.

Given the limited options for direct investigations of the properties
of metal-free stars, indirect studies are the next best strategy. Such
observations have been carried out to constrain the IMF of Pop~III
stars by studying the properties of the most metal-poor stars in the
local universe \citep{freeman02,beers05,frebel07}, where remnants of
the first Population III stars could be found \citep{tss08}. The idea
behind these ``galactic archeology'' campaigns is that stars with very
low metallicities $Z \lesssim 10^{-3.5}~ Z_{\sun}$ (approximately the
critical metallicity marking the transition from Pop~III to Pop~II)
are most likely second-generation objects formed from gas enriched by
only one previous generation of stars. If this first generation of
stars is very massive ($M \gtrsim 100~ \msun$), \citet{heger03} find
that a significant fraction with initial masses of $140-260~\msun$
will explode as pair instability supernovae (PISN) and produce no
r-process elements \citep{heger02,tumlinson04}. Surveys for extremely
metal-poor stars in the Milky Way (MW) have found hundreds of stars
with metallicity $Z \lesssim 10^{-3} Z_{\sun}$
\citep{beers05,schoerck2008}, but so far none of those stars matches
the nucleosynthetic pattern of PISN progenitors
\citep{tumlinson06,frebel09}. Based on this observational evidence, it
has been proposed that the IMF of Population III stars is peaked at
lower masses, $\sim 30~\msun$ \citep{tumlinson06}. This is in conflict
with the results from numerical simulations of Pop~III star-formation,
which suggest higher masses for metal-free stars.

Several past investigations studied the connection between observed
metallicity distribution of Galactic metal-poor stars and Population
III stars. For example, \citet{karlsson08} follow in detail the
self-enrichment of a single $\sim 10^8~M_{\sun}$ halo and argue that
second generation stars enriched by PISNe would reach a metallicity
[Ca/H] $\sim -2.5$, higher than the cutoff used for current EMP
searches. Hence, the lack of PISNe signatures in stars considered to
be second generation could simply be a result of an observational
bias. \citet{karlsson08} show that their results are insensitive to a
redshift-independent change of the PopIII star formation efficiency.
However their work does not take into account the suppression of
Population III star formation in minihalos due to LW feedback, which
varies with redshift. Therefore the metallicity overshooting by PISNe
could possibly be less severe. A different explanation has been
suggested by \citet{salvadori07}. They use an analytic merger-tree
code aimed at reproducing the $z=0$ Galactic metallicity distribution.
As a result of fine tuning their free parameters to match the $z=0$
observations, they find a negligible probability of observing second
generation stars. The robustness of their conclusion is difficult to
evaluate, especially because their model has an idealized treatment of
metal outflows and assumes a one-zone instantaneous mixing of metals
for the gas outside star-forming halos. In addition, Population III
stars are only assumed to form in halos with $T_{\rm vir}\geq 10^4~K$,
completely neglecting formation in minihalos cooled by $H_2$. Similar
conclusions to \citet{salvadori07} have also been reached by
\citet{komiya07,komiya09} who also use an analytical merger-tree code
(with instantaneous mixing approximation) and without taking into
account the effects of radiative feedback on Population III star
formation.

Both of these physical effects are examined in this paper, with the   
goal of quantifying the IGM enrichment by Population~III stars and 
predicting the expected fraction of EMP stars with PISNe signatures.
The context of our study takes into account the complex interplay of
radiative and chemical feedback during the reionization epoch. We
employ high-resolution cosmological simulations of structure formation,
both for an average region of the Universe (in a comoving volume of 
$10^3$~Mpc$^3$) and for three dark-matter halos similar in mass to the 
Milky Way. We post-process these simulations by taking into account 
additional physical processes that regulate star formation, beyond those 
considered in past investigations such as 
\citet{tumlinson06,salvadori07,komiya07,komiya09,karlsson08}.  These
processes include:
(1) radiative feedback on the cooling of metal-free minihalos; 
(2) spatial distribution of metal outflows from protogalaxies, 
    missing in investigations based on analytical merger-tree codes; and 
(3) impact of the cosmic microwave background (CMB) temperature on the 
    typical mass of Population~II stars formed at high-redshift
\citep{clarke03,tumlinson07,smith_b09}.

This paper is organized as follows. In Section~\ref{sec:transition} we
introduce our model and discuss the transition from Population~III to
Population~II star formation and investigate its consequences for the
chemical enrichment of second-generation gas in Section~\ref{sec:av_box}. 
Section~\ref{sec:av_box} is based on analytical modeling and cosmological 
simulations representative of an average region of the high-redshift Universe. 
In Section~\ref{sec:origins} we present results from three
constrained-realization simulations of halos like the Milky Way, and we 
discuss the differences with the averaged results obtained for a
random region of the Universe. Section~\ref{sec:conc} concludes with
a discussion of Milky Way EMP surveys needed to probe the metallicity
distributions and set limits on PISN progenitors at $z \leq 10$.

\section{Transition from Population III to Population II}\label{sec:transition}

In this analysis, we adopt the star formation model described in
\citet{ts09} and \citet{trenti09b}. We consider the halo dynamics in a
dark-matter scenario, based either on Press-Schechter modeling plus an
analytical expression for self-enrichment probability (see
\citealt{ts07a}) or on detailed halo-merger histories derived from
cosmological simulations \citep{trenti09b}. Dark matter halos are
populated according to an analytical cooling model that takes into
account radiative feedback in the LW bands of H$_2$, self-enrichment, 
and wind enrichment, the latter limited to the star formation histories from
numerical simulations. A single, massive, Pop~III star is assumed to
form in halos with $Z < Z_{\rm crit}$ and $10^3 \mathrm{K} \leqslant
T_{\rm vir} \leqslant 10^4 \mathrm{K}$. The precise value of $T_{\rm vir}$ 
required for Pop~III formation depends on redshift and is derived considering 
the negative (H$_2$ LW-band) feedback self-consistently
determined from the star formation rate. Pop~III stars are assumed to
end their life as a Pair Instability Supernova \citep{heger02}, while a
constant fraction of gas (typically $f_* \approx 10^{-2}$) is converted into 
stars for metal-enriched halos with a minimum $T_{\rm vir} \geq 10^4$ K. 
Multiple Pop~III stars are optionally allowed to form in the same halo when 
Ly$\alpha$ cooling is possible, in metal-free halos with 
$T_{\rm vir} \geq 10^4$~K. 

Because of the large amount of metals (50--100~$M_{\sun}$) released by a 
PISN explosion \citep{heger02,heger03,galyam09}, a single Population~III 
progenitor is sufficient to enrich a descendant halo with 
$T_{\rm vir}\sim 10^4$ K to $Z_{\rm crit}$. The violent relaxation 
\citep{lyndenbell67} associated with the virialization of the descendant
halo provides efficient mixing of the metals, initially concentrated
in a $\sim 1~\mathrm{kpc}$ region around the site of explosion
\citep{bromm03}.  However, the metal outflow from a PISN is confined to
the region occupied by a $T_{\rm vir}\lesssim 10^4$~K halo \citep{bromm03}.
Hence, Population~III stars formed in isolation only contribute to genetic 
enrichment of their descendant halos and are not expected to pollute nearby 
halos. Metal outflows are instead expected from the Population~II galaxies 
in halos with $T_{\rm vir} \gtrsim 10^4$~K. These outflows are estimated to 
travel at average velocities of $40-100~\mathrm{km~s^{-1}}$ and can enrich
a region of radius $\sim 100~h^{-1}~\mathrm{kpc}$ 
\citep{madau01,furlanetto03,tumlinson04}. We assume conservatively a 
star-formation efficiency $f_* = 0.01$ and a metal yield of $y_m = 1/40$,
derived for a local $z=0$ IMF, hence likely an underestimate at $z\gtrsim 6$ 
given the effect of the CMB temperature.  We find that an outflow originating 
from a $M=10^{8}~M_{\sun}$ halo can enrich to $Z\sim Z_{\rm crit}$ a fully
virialized $10^8~M_{\sun}$ halo within $d\lesssim 33~h^{-1}~\mathrm{kpc}$ or 
a region containing the same mass in the process of virialization 
(with overdensity $\Delta \rho / \langle \rho \rangle = 2$) out to 
$d \lesssim 150~h^{-1}~\mathrm{kpc}$. To model wind enrichment from the 
cosmological simulations, we therefore save snapshots with high-frequency 
($\Delta z = 0.125$ at $z\leqslant 10$ and $\Delta z = 0.25$ at 
$10 \leqslant z\leqslant 15 $) and track during post-processing the expansion 
of metal outflows originating from Population~II halos. We assume that a halo 
with $T_{\rm vir}\gtrsim 10^4$ K is enriched to $Z\gtrsim Z_{\rm crit}$ 
(if not already above this threshold) once it is reached by the front of a 
metal outflow.

Therefore, in our model, the gas is either pristine or above the
critical metallicity. For a lower value of $Z_{\rm crit}$, possible in
presence of dust \citep{schneider06}, this assumption would be even
better justified. Our standard analysis is carried out for
$v_{wind} = 60~\mathrm{km~s^{-1}}$, but we also investigate
$v_{wind} = 40~\mathrm{km~s^{-1}}$ and $v_{wind} = 100~\mathrm{km~s^{-1}}$. 
Note that in our treatment of metal enrichment and star formation,
metal-free halos with $T_{\rm vir} \geqslant 10^4$ K are assumed to form 
a first generation of massive metal-free stars, immediately followed by
a second generation of stars with $Z\gtrsim Z_{\rm crit}$. This implies
that we are effectively setting the lifetime of a Population~III star
to zero and assuming instantaneous recycling for genetic enrichment in
$T_{vir} \geqslant 10^4$ K halos. In this approximation, we neglect 
the small fraction of halos that are quickly reached by a metal outflow,
within a few Myr after hosting a burst of Population~III star formation. 
Because winds propagate for $\gtrsim 100$ Myr before reaching a metal-free 
halo, accounting for the lifetime of a massive Population~III star would be 
equivalent to increasing $v_{wind}$ by $\lesssim 5\%$.  This has a minimal 
impact in our model given the larger uncertainty on other parameters.

We now summarize the main results of our model \citep{trenti09b},
crucial to interpret the observed enrichment pattern in Galactic
extremely metal-poor stars: \begin{enumerate}

\item Population III star formation in minihalos cooled by H$_2$
  starts at very high redshift ($z\gtrsim 45$) for a comoving volume 
  comparable to that of a MW-like halo. Radiative feedback begins
  quenching this mode of star-formation at $z\sim 35$ and completely 
  suppresses it by $z \sim 15$. For a given star formation rate, the
  flux in the LW bands is almost IMF and metallicity independent, a
  very marked difference compared to the ionizing flux, which
  increases significantly for a top-heavy IMF and for extremely low or
  zero metallicity (\citealt{schaerer03}, Table 4). 

\item The metal-enriched star formation rate becomes higher than the
  metal-free rate at $z \sim 25$ and continues to increase almost 
  exponentially until $z\sim 10$. At $z\sim 20$, Pop~ II stars are 
  already the dominant star-formation mode and implies that 
  metal-free stars are probably minor players in reionizing the
  Universe. At $z\lesssim 25$, LW photons are primarily produced by
  metal enriched stars. 

\item Population III star-formation continues to be possible at $z < 15$ 
    at a comoving rate $\sim 10^{-6}~\msun~\mathrm{yr^{-1}~Mpc^{-3}}$ in 
    larger halos with $T_{\rm vir} \gtrsim 10^4$~K, insensitive to the 
    LW photo-dissociating background. Winds from nearby Pop~II halos 
    enrich all the IGM by $z \sim 4 $, progressively reducing the 
    Pop~III star formation rate, beginning at $z \sim 10$.

\end{enumerate}

\newpage

\section{Second-generation Star Formation in average regions }\label{sec:av_box}

The complex star-formation history derived in \citet{trenti09b} raises
two questions in the context of the observed chemical abundances of
EMP stars: First, what are the expected chemical progenitors of EMP stars? 
Are they Pop~III stars in minihalos, Pop~III stars in the larger $T_{\rm vir}
\sim10^4$ K halos (via self-enrichment), or Pop~II stars (via winds)?
Second, over which redshift range did the \emph{currently observable} 
EMP stars form?

We first address these questions for an average region of the
Universe, using the $N=1024^3$ dark matter particle cosmological
simulation of \citet{trenti09b}, with a comoving volume
$V=10^3~\mathrm{Mpc^3}$, a single-particle mass $3.4\times 10^4~M_{\sun}$,
and a force resolution of $0.16~h^{-1}$~kpc. We adopt a
cosmology based on the fifth-year WMAP data \citep{komatsu08}:
$\Omega_{\Lambda}=0.72$, $\Omega_{m}=0.28$, $\Omega_{b}=0.0462$,
$H_0=70~$ km~s$^{-1}$~Mpc$^{-1}$, and $\sigma_8=0.817$. The initial
conditions have been generated at $z=199$ with a code based on the
Grafic algorithm \citep{bertschinger01} using a $\Lambda CDM$ transfer
function computed via the fit by \citet{eisenstein99} with spectral
index $n_s=0.96$. Halos are identified with a friend-of-friend halo
finder \citep{davis85} using a linking length equal to 0.2 of the mean
particle separation. A halo merger tree is constructed from the
particle IDs to track metal enrichment derived from progenitor halos.

We define ``second-generation gas'' as gas at $Z>Z_{\rm crit}$, residing
in a $T_{\rm vir} \geqslant 10^4$ K halo with no progenitor that 
experienced Population II star formation in a previous snapshot. Next,
we must distinguish between ``second-generation gas" enriched by
Pop~II or Pop~III stars. The earliest gas was enriched by Pop~III
stars in dark matter halos with $T_{\rm vir} \geq 10^4$~K. During the
halo merging history, such halos had at least one progenitor halo that
hosted a Pop~III star, but no progenitors with Pop~II stars. In
contrast, second-generation gas enriched by Pop~II stars is defined as
gas in a dark matter halo with $T_{\rm vir} \geq 10^4$~K whose
progenitor halos had no star formation and were chemically enriched by
winds originating in a nearby halo\footnote{The majority of halos with
$T_{\rm vir}>10^4$~K in the simulation box have metal-enriched halos as 
progenitors and thus host at least third-generation stars.}.

\begin{figure} 
  \plotone{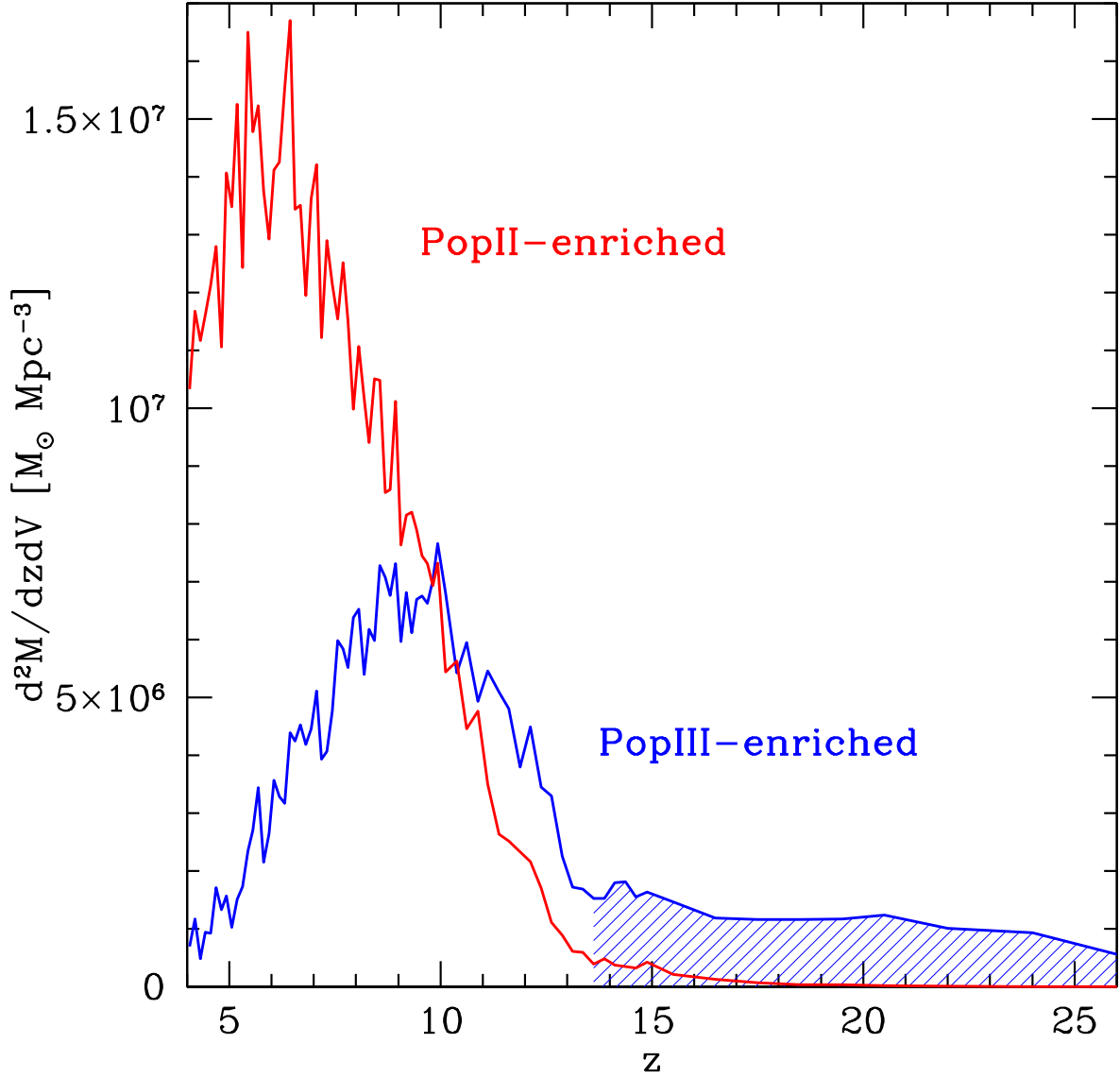}\caption{Formation rate per unit redshift
   of second-generation gas ($\msun~\mathrm{Mpc^{-3}}~dz^{-1}$) enriched 
   to metallicity $Z \sim 10^{-3.5}~Z_{\sun}$ as measured from a 
   $10^3~ \mathrm{Mpc^3}$ cosmological simulation that takes into account 
   radiative feedback for Population III formation, self-enrichment of 
   halos, and metal winds propagating at $60~\mathrm{km~s^{-1}}$ (see 
   \citealt{trenti09b}).
   The rate of gas enriched by Pop~III stars is shown as a
   solid blue line, while the rate for Pop~II-enriched gas is
   shown as solid red line. The blue shaded area represents the gas
   enriched by Population III stars formed in minihalos (cooled by
   H$_2$). Most second-generation stars in the volume are
   formed near the end of the reionization era, at 
   redshifts $z \lesssim 10$. }
 \label{fig:second_gen} \end{figure}

Figure~\ref{fig:second_gen} shows that most (about $60\%$) of the
second-generation gas is enriched by Pop~II winds. In addition,
Pop~III stars in minihalos (formed at $z \gtrsim 14$) only contribute
to a small fraction (about $10\%$) of second-generation enrichment;
this gas is highlighted as the shaded area in
Fig.~\ref{fig:second_gen}. This surprising result can be understood
from the star-formation rate history plotted in Fig.~1 of
\citet{ts09}. The Pop~III star formation rate per unit time remains
relatively constant from $z\sim 40$ to $z\sim 10$, so that most of the
stars are formed at the lower end of the interval. (Note that
cosmological time scales as $t \propto (1+z)^{-3/2}$). As time passes,
metal-enriched winds from Pop~II galaxies fill progressively more
volume, enriching pockets of low-metallicity gas that never underwent
self-enrichment because the strong radiative (LW) background
suppressed H$_2$ cooling (see also Fig.~4 in \citealt{trenti09b}).
Figure~\ref{fig:second_gen} was obtained for $v_{\rm
  wind}=60~\mathrm{km~s^{-1}}$. Faster winds increase the fraction of
the gas enriched by Pop~II stars, while slower winds favor a higher
fraction of Pop~III IGM enrichment (see Fig.~5 in
\citealt{trenti09b}).

To translate the fraction of second-generation gas enriched by Pop~III
stars into a prediction for EMP star surveys, we need to address the
second question raised above, namely when did \emph{currently
  observable} EMP stars form? These stars can be observed only if they
are bright nearby main-sequence stars or lie along the giant branch.
Given that second-generation stars are quite old (they formed at
$z\gtrsim 4$ according to Fig.~\ref{fig:second_gen}), we can observe them 
only if their initial mass was $M \lesssim 0.9~ \msun$. For the present-day
IMF, most of the stars formed lie below this cut-off. However, this
was not necessarily true at earlier times, because coupling with the
CMB radiation prevents the gas to cool below $T_{CMB}$
\citep{clarke03,tumlinson07,smith_b09}\footnote{Note that the coupling 
is relevant only for gas at $Z\gtrsim Z_{\rm crit}$, because Population~III 
halos cannot easily reach temperatures below $T\sim 200$ K, already higher 
than $T_{CMB}$ at $z<70$.}. In fact, the fragmentation scale of protostellar 
clouds is probably set by the Jeans mass \citep{larson05,tumlinson07}, which
depends on the temperature $T_{\rm
  min}$ and density $\rho$ of the gas:
%%%%%%%
\be
M_J = \left ( \frac{\pi k T_{min}}{2 m_H G} \right ) ^{3/2} \rho^{-1/2}.
\ee
%%%%%%%
By adopting a power-law equation of state, \citet{tumlinson07}
expresses the typical stellar mass in the IMF as a function of redshift:
%%%%%%%%%%%%%
\be \label{eq:mc}
M_C =   M_{\rm norm}  \left \{ \frac{\mathrm{max}[2.73(1+z),8]}{10~ \mathrm{K}} \right \} ^{\alpha},
\ee
%%%%%%%%%%%
where $\alpha$ depends on the equation of state, with an expected range 
$1.7 \leq \alpha \leq 3.35$, and the normalization $M_{\rm norm}$ is defined as:
%%%%
\be
M_{\rm norm} = \frac{0.5~\msun}{0.8^{\alpha}}  \; ,
\ee
%%%%
based on a typical temperature of $8$ K and a typical mass of
$0.5~\msun$ observed in the local Universe \citep{kroupa2002}.
Equation~\ref{eq:mc} can then be used to define the IMF, parameterized in 
log-normal form (Eq.~4 in \citealt{tumlinson07}):
%%%%%
\be \label{eq:imf}
\ln{\left( \frac{dN}{d \ln M} \right)} = A - \frac{1}{2\sigma^2}
   \left[  \ln{ \left( \frac{M}{M_C} \right)} \right]^2,
\ee
%%%%%
where $A$ is a normalization constant and $\sigma$ is a free parameter
of order unity that controls the width of the distribution.

\begin{figure} 
 \includegraphics[scale=0.44]{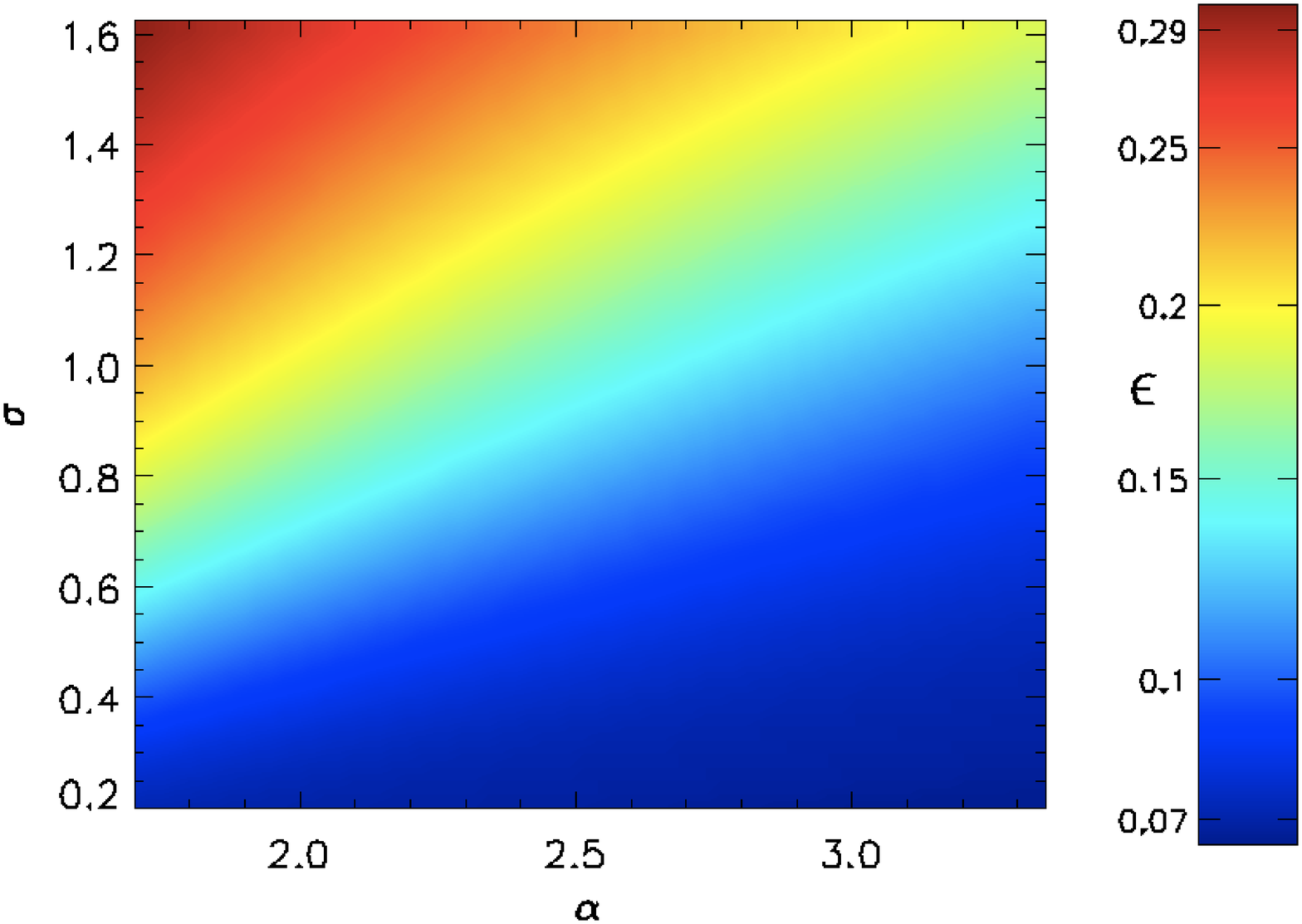}  
 \includegraphics[scale=0.4]{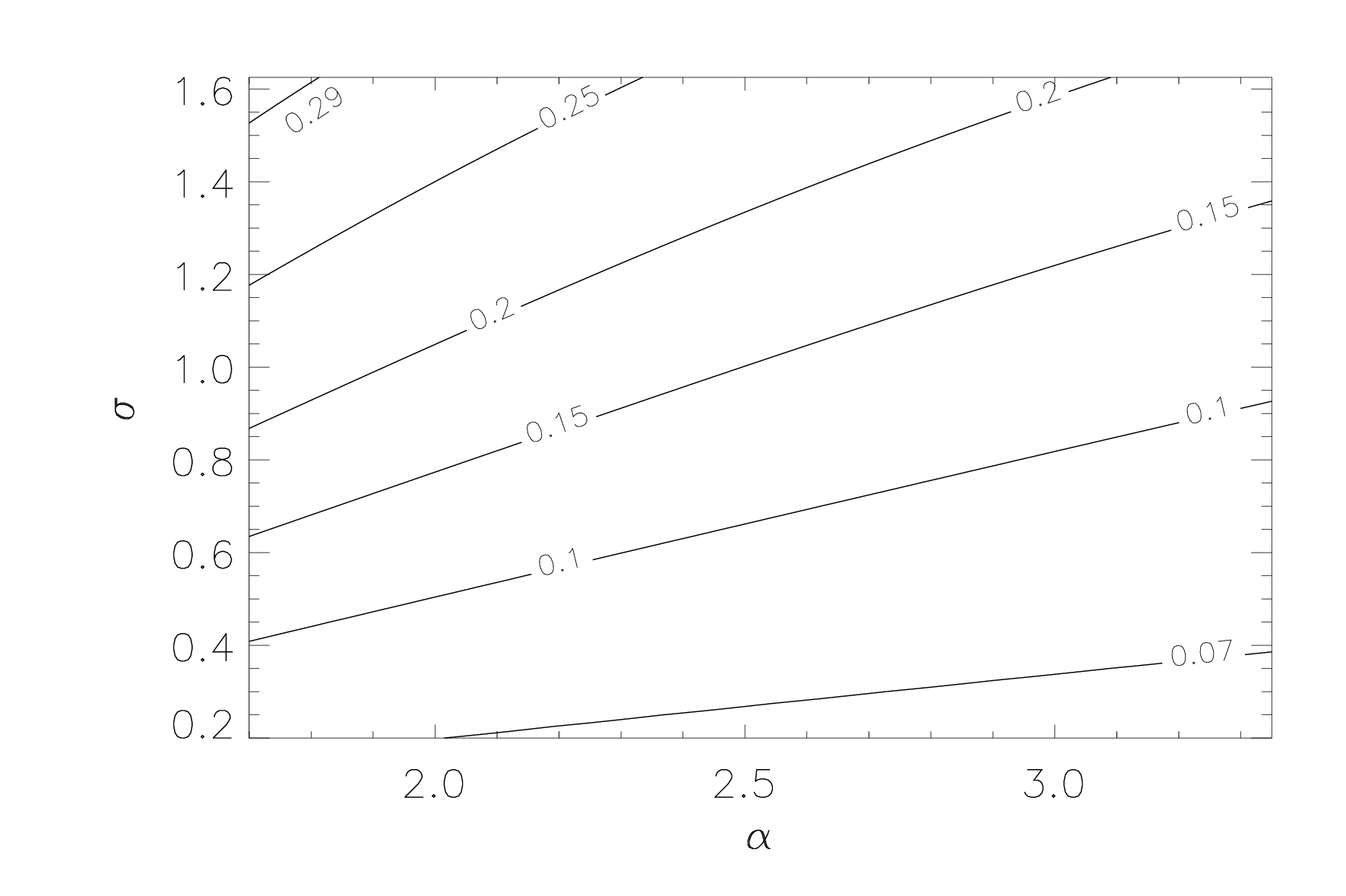}  
 \caption{Upper panel: Image showing the ratio $\epsilon$ of stars
   enriched by Population~III to total number of second-generation
   stars  (with initial mass $M \leq 0.9 ~\msun$) on the main sequence 
   or giant branch in the current universe, plotted as function of
   log-normal IMF parameters $\alpha$ and $\sigma$. Color coding for
   $\epsilon$ is given by the color bar on the right. Lower panel:
   contour lines for
   $\epsilon(\alpha,\sigma)=0.07,0.1,0.15,0.20,0.25,0.29$ (from
   bottom to top). }
\label{fig:eps_std} 
\end{figure}

From Eq.~\ref{eq:mc} and Eq.~\ref{eq:imf} we can compute the number
of second-generation stars (defined as stars formed out of
``second-generation'' gas) still observable today, with $M \leq
0.9~\msun$. If we are interested only in the relative fraction of EMP
enriched by Pop~III vs.\ those enriched by Pop~II, our
result does not depend on the EMP-star formation efficiency, as long
as we assume a constant efficiency per unit mass in transforming gas
to stars. Figure~\ref{fig:eps_std} shows this fraction, defined as
\begin{equation}
   \epsilon = N_{\rm PopIII-enriched} / 
     (N_{\rm PopII-enriched} + N_{\rm PopIII-enriched})
\end{equation}
as a function of ($\sigma$, $\alpha$). Here, $\sigma$ is the IMF
mass width , and $\alpha$ is the exponent that regulates the typical
stellar mass as a function of the CMB temperature (Eq.~\ref{eq:mc}).
Even for a very broad IMF ($\sigma = 1.4$) and a shallow redshift
evolution of the typical stellar mass ($\alpha=1.7$), only up to $\sim
25\%$ of observable EMP stars are formed out of Pop~III-enriched gas.
For intermediate values of the IMF parameters, we obtain $\epsilon
\sim 0.15$, while an extremely narrow IMF reduces $\epsilon$ to $\sim
0.06$. These numbers imply that a significant sample of EMP stars with
metallicity $Z \sim 10^{-3.5} Z_{\sun}$ is required to rule out the
null hypothesis that Pop~III stars are very massive and preferentially
explode as PISN (see Table~\ref{tab:sample_size}). For a Poisson
distribution, $P(k) = N_{exp}^k \exp{(-N_{exp})} /k!$, with expected
value $N_{exp}$, the probability of zero events is $P(0)=
\exp{(-N_{exp})}$. Thus, the minimum second-generation sample size to
rule out that a fraction $\eta$ of Population III stars exploded as
PISNe at a confidence level $\xi$  ($0 < \xi < 1$) is:
%%%%%%%%
\be\label{eq:min_n}
N_{min} = \frac{-\log{(1-\xi)}}{\epsilon(\alpha,\sigma) \eta}.
\ee
%%%%%%%%
For example, to rule out at 99\% of confidence level ($\xi=0.99$) that 
50\% of Pop~III stars explode as PISN, we need 41--135 second-generation 
stars, depending on the parameters assumed for the IMF of the EMP stars.

\begin{figure} 
\includegraphics[scale=0.44]{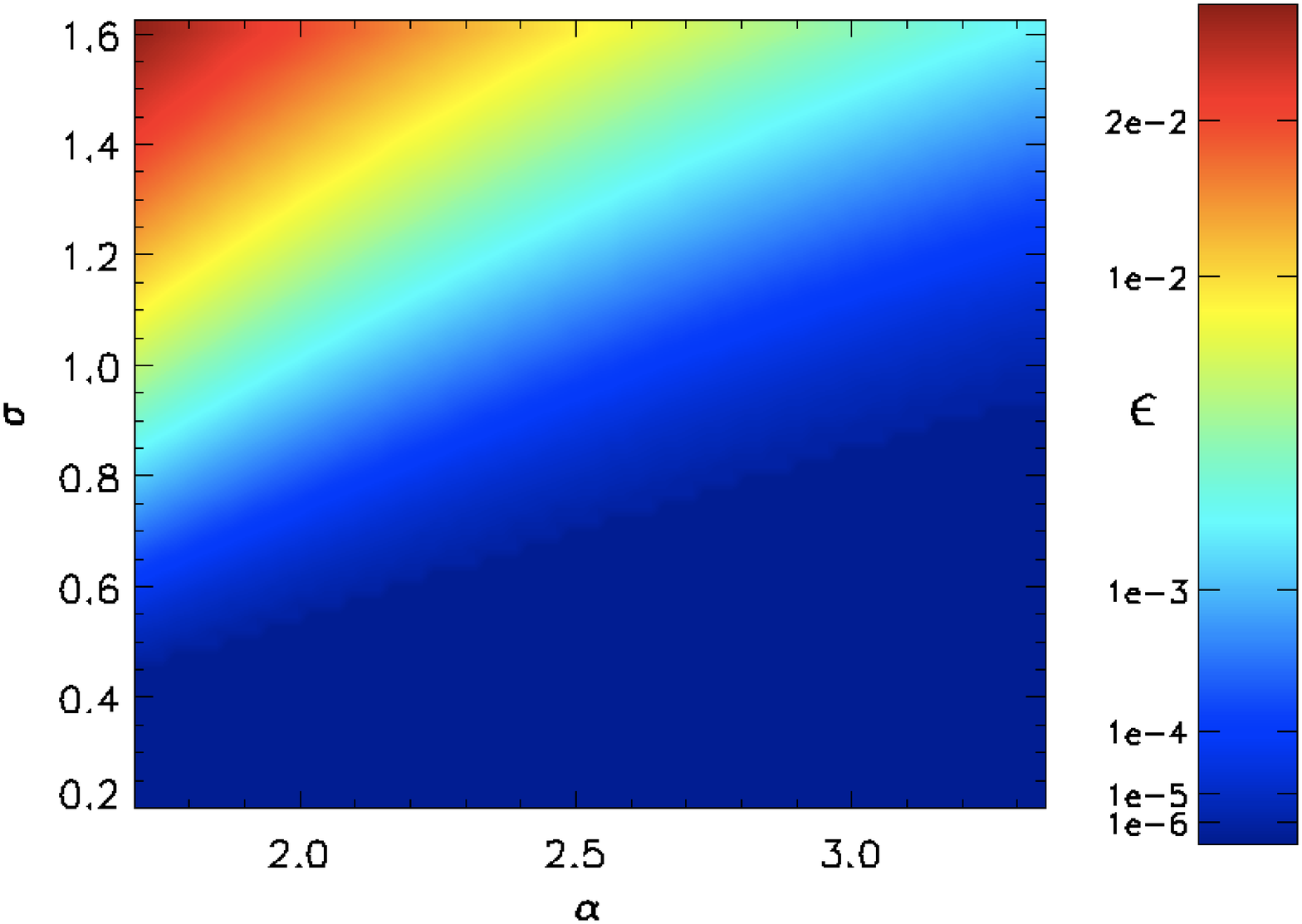}
 \includegraphics[scale=0.4]{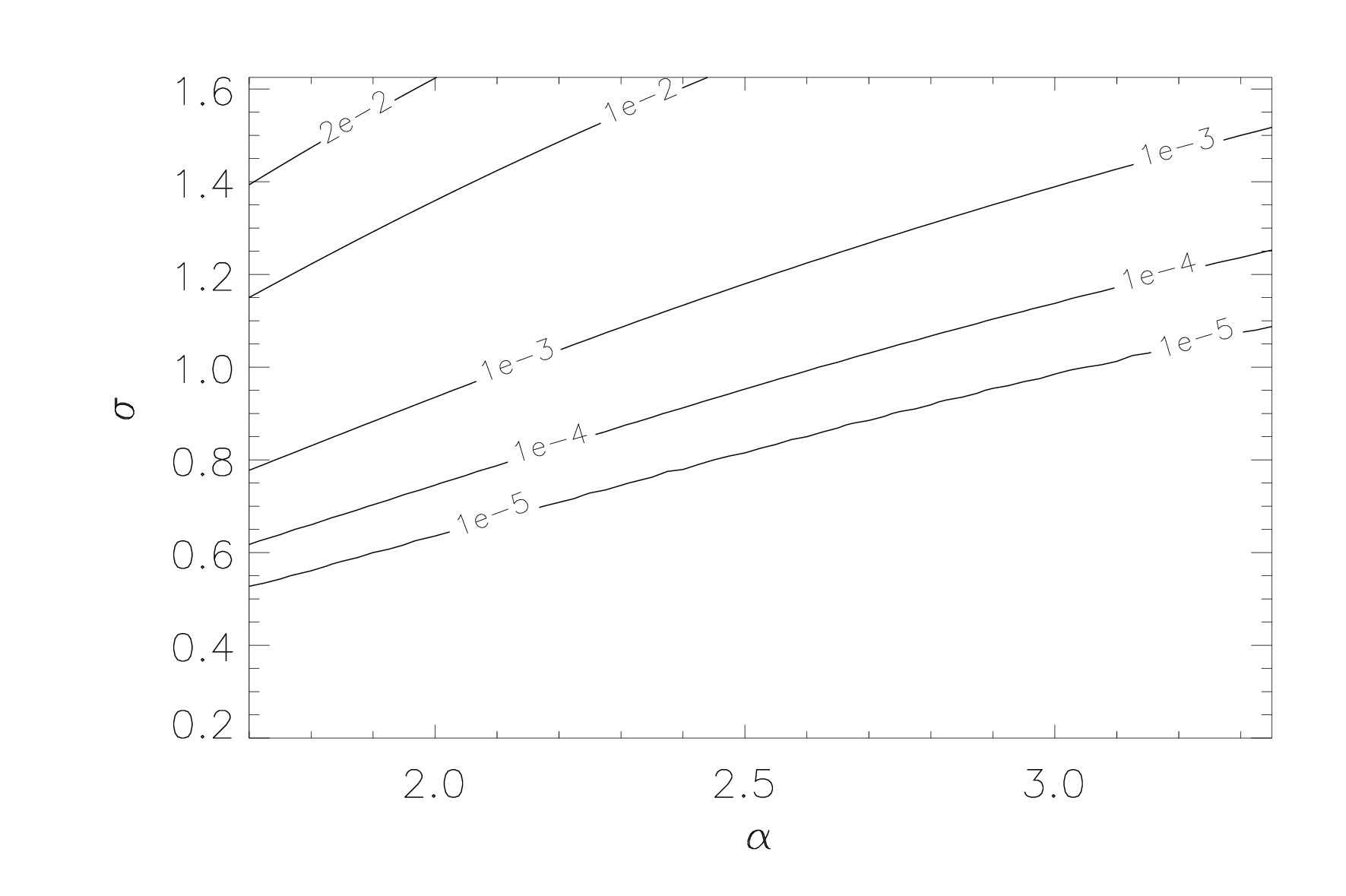}
 \caption{Same as Fig.~\ref{fig:eps_std}, but considering only 
  second-generation stars enriched by the Population~III stars in minihalos.
  This second-generation gas is represented as shaded area in
  Fig.~\ref{fig:second_gen}.  Contour lines are for much smaller values of 
   $\epsilon = 10^{-5}, 10^{-4}, 10^{-3},10^{-2},2\times10^{-2}$ from bottom to top. 
   Studies of extremely metal-poor stars will need a very large 
   sample to constrain the IMF of Population~III stars progenitors in 
   minihalos.}

\label{fig:eps_minihalos} 
\end{figure}

With the same approach, we can estimate the size of the sample
required to investigate the IMF of Pop~III stars in minihalos at $z
\gtrsim 15$. The fraction of early-time Pop~III-enriched
second-generation stars for this case is shown in
Fig.~\ref{fig:eps_minihalos}. We note the limited amount of gas
enriched by this sub-class of Pop~III stars (see
Fig.~\ref{fig:second_gen}) as well as the increase of the typical EMP
stellar mass as a function of redshift (see Eq.~\ref{eq:mc}). It is
highly unlikely to observe one of these second-generation stars today.
In fact, we find $\epsilon \leq 0.02$ over the entire EMP-IMF
parameter space. The main controlling parameter is $\sigma$, the IMF
width. Under the most optimistic assumptions for $\sigma$ and $\alpha$
in Table~\ref{tab:sample_size}, about 1775 second-generation stars are
required to reject the hypothesis of 50\% PISN from metal-free stars
in minihalos at 99\% CL. If the EMP-IMF is narrower, a sample size
above $10^4$ sources is required. As we discuss in \citet{trenti09b},
the formation rate of minihalos, as measured in our cosmological
simulation, might be slightly underestimated owing to resolution
issues. If this is the case, the size of the second-generation star
sample is reduced accordingly. Conservatively assuming a factor-of-two
higher Pop~III formation rate in minihalos implies a sample size
$\gtrsim 900$ second-generation stars.

\section{Second-Generation Star Formation in Milky-Way like halos}\label{sec:origins}

The star formation model in Section~\ref{sec:av_box} describes an
average region of the Universe at $z \geq 4$. Instead, the MW
progenitors lived in a region with an above-average halo formation
rate because of the presence of the large-scale overdensity that
created the MW and the Local Group at $z \lesssim 1$. As a result, the
star-formation history of the region is enhanced and tends to be
shifted earlier in time compared with a typical region of the
Universe. This introduces two competing factors that can affect the
expected fraction of observable second generation stars enriched by
Pop~III stars. First, metal winds have had less time to reach nearby halos,
so we expect a larger fraction of second generation gas enriched by
Pop~III stars.  This effect is balanced by the higher CMB temperature floor
present at earlier times, which makes it more difficult for stars
formed out of this gas to have sub-solar masses and thus be observable
today.

To quantify this scenario and to obtain a detailed prediction to
compare with EMP surveys in the MW, we consider a new set of
cosmological simulations, with the same setup of the one considered in
Section~\ref{sec:av_box}, except for the initial conditions. Here, we
use the Grafic package \citep{bertschinger01} and set the initial
conditions to host a MW-like halo at $z=0$ at the center of the
simulation box. Because of the prohibitive computational time needed
to evolve to $z=0$ a $V=10^3~\mathrm{Mpc^3}$ volume with $N=1024^3$
particles, we first run a low-resolution ($N_{lr}=128^3$) version of
the simulation box to $z=0$ to measure directly the central halo mass
realized in the initial conditions. We consider three different
realizations of central halos to explore a range of masses and
formation histories: $M_1=3.15\times 10^{12}~M_{\sun}$ (corresponding
to $172,981$ low-resolution particles), $M_2 =2.60\times 10^{12}~M_{\sun}$,
and $M_3=1.59\times 10^{12}~M_{\sun}$. From $z=1$
to $z=0$, halo 1 acquires 45\% of its mass, similar to halo 3 (42\%),
while halo 2 experiences a major merger and acquires 59\% of its mass.
A projection of the halos at $z=0$ (region size $\sim 1~\mathrm{Mpc^2}$)
from the low-resolution run is shown in the upper left panel of
Figs.~\ref{fig:halo0}-\ref{fig:halo2}.

We then create the final full resolution ($N=1024^3$) box by keeping
the same large-scale structure of the low-resolution version using the
\citet{hoffman91} method. Each low-resolution particle is thus
associated with $n_c=1024^3/128^3=512$ ``child'' particles in the
high-resolution run. Based on the initial positions, we map the IDs of
the particles in the central halo of the low resolution realization
into the set of high-resolution particle IDs expected to be part of
this central halo. We tested this method extensively with a series of
constrained simulations at different resolution and found that, for a
halo resolved with $\sim 10^5$ low-resolution particles, the mass (and
hence halo membership) is accurate at the $\gtrsim 97\%$ level (Trenti et
al., submitted). The setup we choose is similar in terms of mass
resolution and number of particles to that adopted by
\citet{tumlinson09}\footnote{The
  preprint was posted on arxiv while this paper was being revised},
with the only difference being that \citet{tumlinson09} binned
particle positions and velocities to reduce the resolution and
complete the run from $z=4$ to $z=0$ rather than using
the \citet{hoffman91} method. 

The post-processing analysis is the same discussed in
Section~\ref{sec:av_box}, with the difference that the results on
second-generation gas are given only for halos that are progenitors of
the central $z=0$ MW-like halo. The comoving volume considered for
normalization is that occupied at $z=199$ by the particles that end up
at $z=0$ in the central halo. The number of particles in this volume 
range from $>88\times 10^6$ for halo 1 to $>44 \times 10^6$ for halo 3.

Figures~\ref{fig:halo0}-\ref{fig:halo2} summarize the results for
the different halos. The formation rate per unit redshift per unit
volume of second generation gas is shown in the upper right panel of
the figures as solid lines and compared to the box-averaged formation
rate (dotted lines). As expected, the regions where the progenitors of
MW-like halos live have an enhanced halo formation rate and higher
clustering than an average region of the Universe. As a consequence,
metal enrichment by Population III stars is more efficient (by
$\gtrsim 50\%$) at $z \gtrsim 10$. This is consistent with a very
simple estimate based on linear theory: the region hosting a MW-like
halo at $z=0$ has an average overdensity $\Delta \rho / \langle \rho
\rangle \sim 0.14$ at $z=15$. This in turn implies that the comoving
number density of halos with $M = 1.4\times 10^7~M_{\sun}$ (hosting
Population III stars at $z=15$, see \citealt{trenti09b}), derived from
the \citet{st99} mass function, is enhanced from $\sim
5.6~\mathrm{Mpc^{-3}}$ to $\sim 8.9~\mathrm{Mpc^{-3}}$.

Because of the higher clustering, winds are also more efficient at
polluting nearby, non self-enriched halos. At $z \lesssim 10$ the
formation rate of Population III enriched gas becomes comparable to
the average in the Universe. The enhanced formation rate of $T_{\rm vir}
\sim 10^4$ K halos is balanced by a higher fraction of such halos
reached by metal outflows. Consequently, the number density of
wind-enriched halos also increases compared to the box average. By
$z\sim 6$ the number density of wind-enriched halos has returned to
values similar to those of an average region of the universe. This can
be again understood in terms of linear theory. $T_{\rm vir} \sim 10^4$
halos are common at $z \lesssim 6$: based on their rarity, measured in
terms of the Press-Schechter variable $\nu = \delta^2_c/\sigma_{DM}^2(M)$, 
these halos have $\nu \sim 1.38$ on
average in the Universe. The $z=5$ overdensity associated with a MW-like
halo at $z=0$ only decreases this rarity to $\nu \sim 1.14$, and the
number density of such overdensities is comparable: $\sim
8.7~\mathrm{Mpc^{-3}}$ on average versus $\sim 9.4~\mathrm{Mpc^{-3}}$
for the MW-like region. The three different halos we consider have all
a similar enrichment history at $z\geqslant 4$.

The predictions for the fraction $\epsilon$ of second-generation stars
enriched by Population III metals is shown in the bottom panels of
Figures~\ref{fig:halo0}-\ref{fig:halo2} considering all Population
III stars (left panels --- equivalent to Figure~\ref{fig:eps_std}) and
only Population III in minihalos (right panel --- equivalent to
Figure~\ref{fig:eps_minihalos}). Given the two competing effects
discussed above (higher halo formation rate and higher efficiency of
wind pollution), the contour lines of $\epsilon(\sigma,\alpha)$ are
similar to those of the box-average. There is a moderate
tendency to lower values for $\epsilon$, implying that the fraction of
second-generation stars enriched only by Population III metals can be
observed today is smaller than the one estimated for the average of
the Universe. Table~\ref{tab:sample_size} contains the minimum number
of stars required to rule out a 50\% PISNe fraction in Population~III
stars for MW halo 1 based on Equation~\ref{eq:min_n}. Minimum sample
sizes for halo 2 and halo 3 are similar. Depending on the IMF
parameters, $(\alpha,\sigma)$, the sample sizes for all Pop~III
enrichment range from a factor of a few difference with the
box-average (at low $\sigma$) to almost equal to the box average (for
high $\sigma$ and low $\alpha$). For enrichment by minihalos-Pop~III in
MW-like progenitors, the required sample sizes are about two times smaller 
than those derived in a random region (because of the factor two increase in the gas enriched by 
these stars compared to an average region of the Universe).

Our main conclusion, namely the very small fraction of second-generation 
stars enriched by Population~III stars observable today, is
similar to that found by \citet{salvadori07}, although their modeling
assumptions are different from ours. In particular, the instantaneous,
single-zone recycling of metals implies a sharp cut-off of Population
III formation, while in our model we find a long tail of metal-free
star formation in halos with lower than average bias (see Fig.~3 in
\citealt{trenti09b}).

Figure~\ref{fig:halo0_wind} explores the impact of $v_{wind}$ on the
results from our model for halo 1. The outflow velocity is the key
parameter controlling the relative fraction of Population III to
Population II enriched gas. In the absence of winds, only self-enrichment
is possible, hence all second-generation gas would be
enriched by Population III metals. On the other hand, faster
propagation of metal outflows suppresses low-redshift Population III
formation. For example, $v_{wind}=100~\mathrm{km~s^{-1}}$ reduces by
$\gtrsim 75$\% the Population~III star formation at $z \lesssim 8$
compared to the standard $v_{wind}=60~\mathrm{km~s^{-1}}$. Slower
outflows, $v_{wind}=40~\mathrm{km~s^{-1}}$, increase the 
Population~III star formation at low redshift by up to a factor two.
This is reflected in the formation rate of second generation gas,
shown in the upper panels of Figure.~\ref{fig:halo0_wind}. The lower
panels contain the $\epsilon(\sigma,\alpha)$ contour lines for these
models. Naturally, the fraction of observable second-generation stars
increases for lower wind speed, but not dramatically. For example,
$\epsilon \leq 0.23$ in the standard model, $\epsilon \leq 0.41$ for
$v_{wind}=40~\mathrm{km~s^{-1}}$, and  $\epsilon \leq 0.10$ for
$v_{wind}=100~\mathrm{km~s^{-1}}$.

\section{Conclusions and Discussion}\label{sec:conc}

In this paper, we propose a novel solution to the conflict between the
extremely top-heavy Pop~III IMF predicted from cosmological
hydrodynamic simulations and the lack of PISN signatures in the metal
abundances of EMP stars observed in the Milky Way halo. 
Compared to past studies with a similar goal, our model takes into 
account additional physical ingredients relevant for star formation and 
chemical enrichment during the reionization epoch.  The most important 
addition is the inclusion of radiative feedback in the UV Lyman-Werner 
bands of H$_2$, bands, which quenches Population~III star formation in 
minihalos and increases the minimum halo mass required for cooling to 
$T_{\rm vir} \sim 10^4$ K by $z\sim 15$ (see \citealt{ts09}). Without 
this increase, we would have obtained a significantly higher self-enrichment 
efficiency and reduced the importance of metal outflows for enriching 
second-generation gas. We
also base our analysis on high-resolution N-body simulations capable
of resolving star forming minihalos, both in a $10^3~\mathrm{Mpc^3}$
(comoving) cosmological volume, as well as in three regions hosting a
$z=0$ halo similar in mass to that of the MW. These simulations allow
us to quantify metal outflows from proto-galaxies, improving the
instantaneous one-zone assumption in \citet{salvadori07}, based on
Press-Schechter merger trees, as well as the work of
\citet{karlsson08}, focused on self-enrichment of a single $10^8
M_{\sun}$ halo. A third novel key ingredient is the inclusion of the
cosmic microwave background (CMB) coupling with cooling gas. This
introduces a floor in the cooling of metal enriched halos and
affects their initial stellar mass function
\citep{clarke03}. This effect was also not modeled in
\citet{salvadori07} or \citet{karlsson08}.

With our assumptions, we have shown that observations and numerical
simulations can be consistent. In fact, we find that the majority of
observable EMP stars with $Z \sim 10^{-3.5}~Z_{\sun}$ were probably
formed out of gas enriched to this critical metallicity by Pop~II winds 
at $z \lesssim 10$. Pop~III stars formed in minihalos cooled by H$_2$ enrich
only a small fraction, $\epsilon \sim 10^{-4}$ to $10^{-2}$, of the
gas that ends up in EMP stars with $M \leq 0.9~\msun$. The fraction of
observable second-generation stars formed in regions enriched by
late-time Pop~III stars is even higher, ranging from $6\%$ to $25\%$,
depending on the details of the modeling, in particular the IMF width
of the EMP stars.

The minimum sample sizes reported in Table~\ref{tab:sample_size} can
be compared with the metallicity distribution of EMP stars based on
Galactic surveys such as the Hamburg/ESO survey \citep{christlieb2008}
and the SDSS/SEGUE survey \citep{ivezic_SEGUE_2008} that have
identified several hundred metal-poor stars with [Fe/H] $< -3.0$
\citep{schoerck2008}. These surveys show a sharp cutoff at [Fe/H] $<
-3.6$, consistent with the theoretical expectation of $Z_{crit} \sim
10^{-3.5} Z_{\sun}$. To infer the Pop~III IMF, we should ideally use
only those EMP stars close to the critical metallicity for the
transition to Pop~II star formation. This would help to avoid
contamination in the sample by third-or-higher generation stars formed
in gas with a lower than average enrichment history. This issue is
especially important if the lack of PISN chemical signatures in
second-generation stars is used to argue against very massive Pop~III
star progenitors. A single PISN producing $10~M_{\sun}$ of iron can
pollute $10^7~M_{\sun}$ of gas (the typical content of a $T_{vir} \sim
10^4$ K halo) to a metallicity [Fe/H] $\sim -3.2$, assuming
homogeneous mixing. This is a level slightly higher than than the
observed cut-off at [Fe/H] $\sim -3.6$, suggesting that it would be
important to improve instantaneous one-zone metal enrichment
assumptions. Except for this caveat, we thus suggest that the analysis
should be confined to a sample of EMP stars with [Fe/H] $\lesssim
-3.4$ to identify second-generation stars. The HES survey has 14 such
stars \citep{schoerck2008}. A few other EMP with [Fe/H] $\lesssim
-3.5$ also have published metallicities (see \citealt{beers05} for a
recent review). Based on this sample size, we would have expected on
average a few of these stars to show PISN signatures in their
abundances if a significant fraction of late-time Pop~III stars were
in the 140-260~$\msun$ range. A larger sample of EMP stars, likely
available in the near future with extensions and/or follow-ups of the
SEGUE and HES surveys, will be able to set stronger limits on the
fraction of PISN originating at $z \lesssim 10$. Investigating the
shape of the IMF for Pop~III stars in minihalos would require an
increase of about two orders of magnitude in the current sample of EMP
stars.

Since the majority of Pop~III numerical investigations have focused on
Pop~III stars in minihalos, there is no inconsistency with the
observed abundances of EMP stars. The sample size used to search for
PISN signatures is simply too small. The current EMP sample is now
approaching a size that can yield solid constraints on the IMF of
late-time Pop~III stars (see Table~\ref{tab:sample_size}). The absence
of PISN signatures is still lacking a significant confidence level,
but it is consistent with the predictions for Pop~III star formation
in (partially) reionized gas. It points toward a typical IMF mass of
$\sim 30~\msun$ \citep{yoshida06}. 

Our scenario for the formation of second-generation stars is also
consistent with the observed distribution of the [Mg/H] ratio for EMP
stars (see Fig.~1 in \citealt{frebel09}). Magnesium abundances
below the minimum level predicted by PISN nucleosynthesis, 
[Mg/H] $\lesssim -3.2$ (see \citealt{heger02,frebel09}) are naturally
expected if the enrichment of second-generation stars with 
$M \leq 140~ \msun$ is driven by Pop~II winds and/or by Pop~III 
stars formed in (partially) reionized gas.

\begin{figure} 
\includegraphics[scale=0.36]{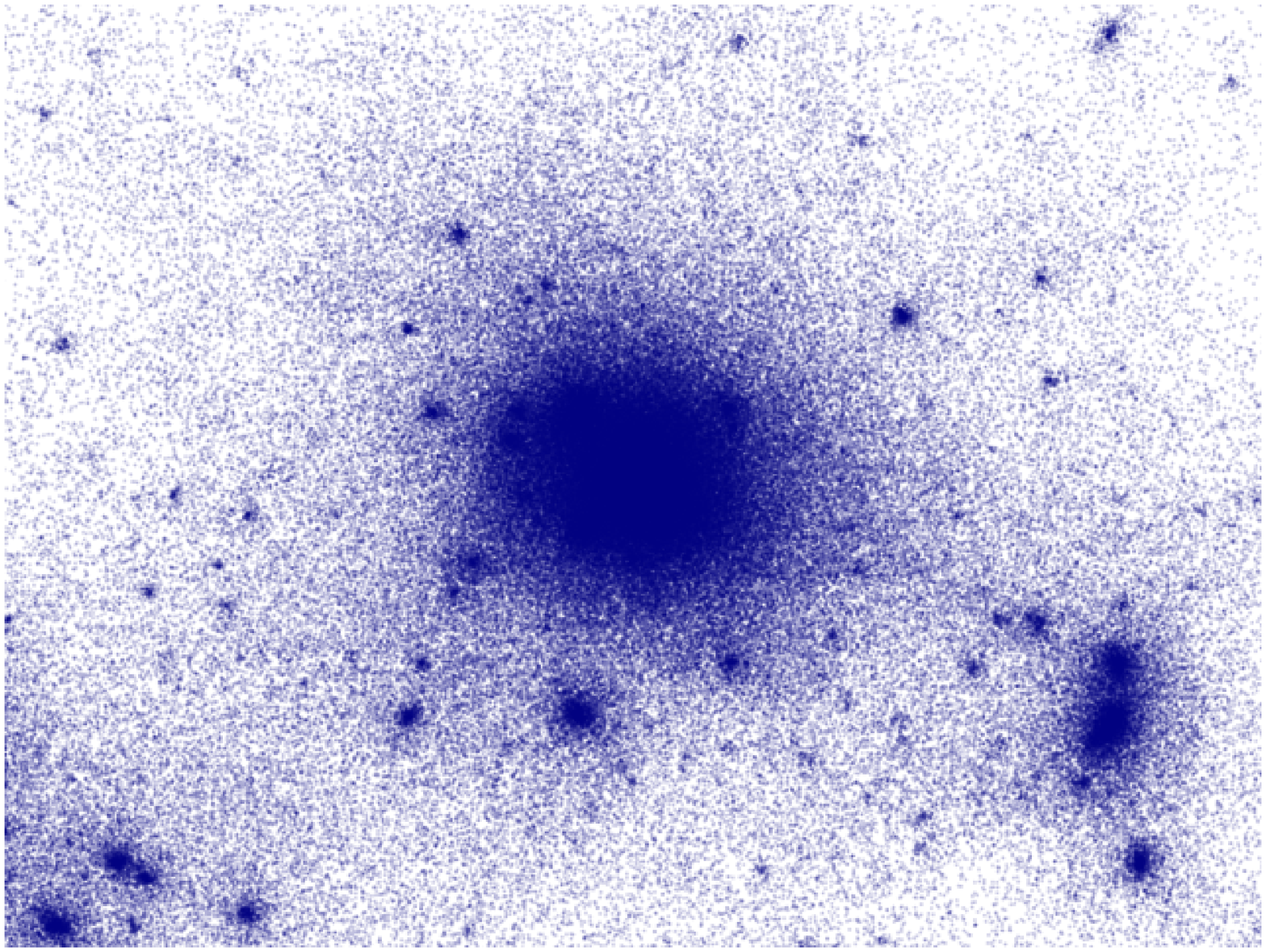}  
\includegraphics[scale=0.38]{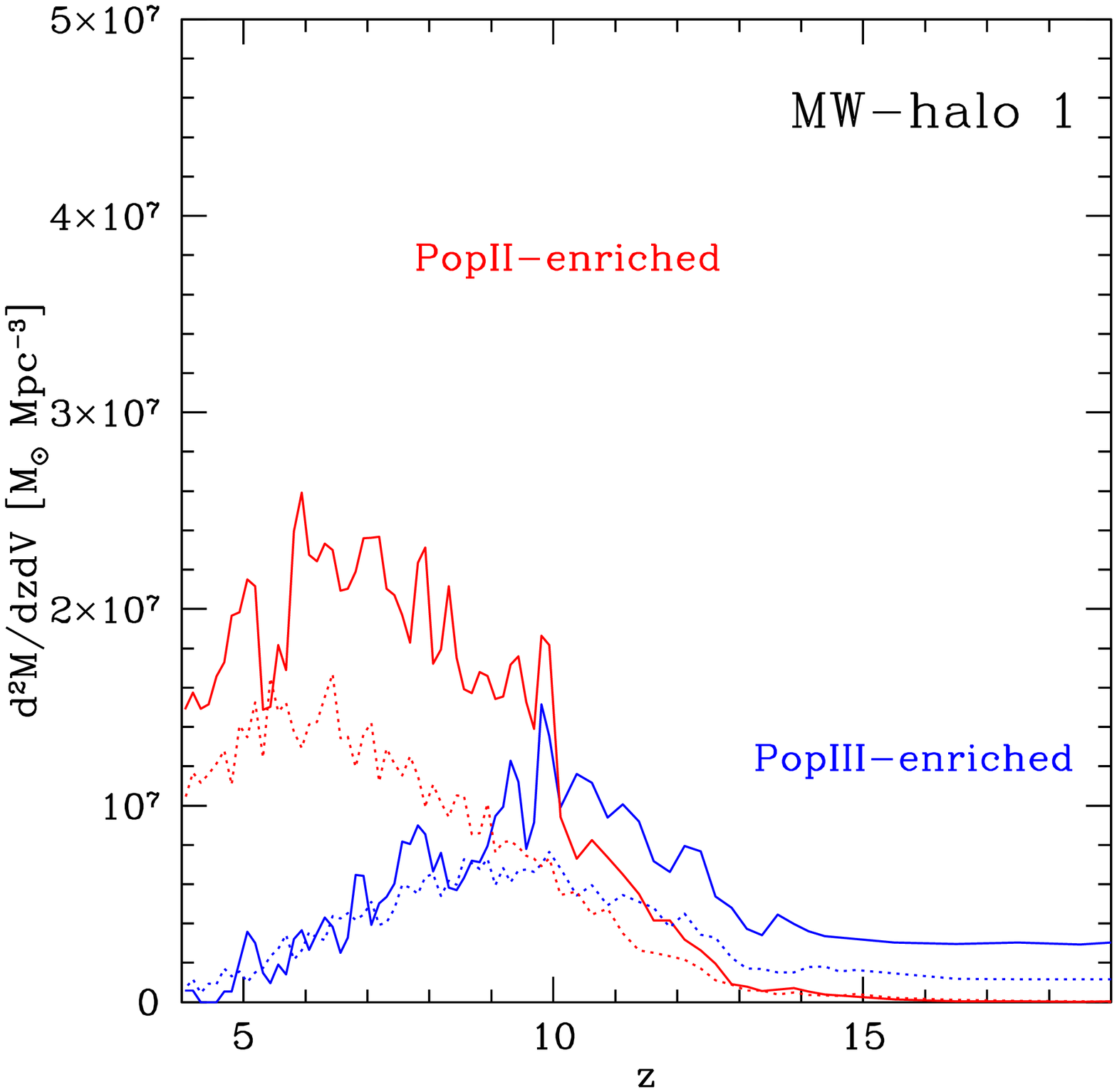}
\includegraphics[scale=0.28]{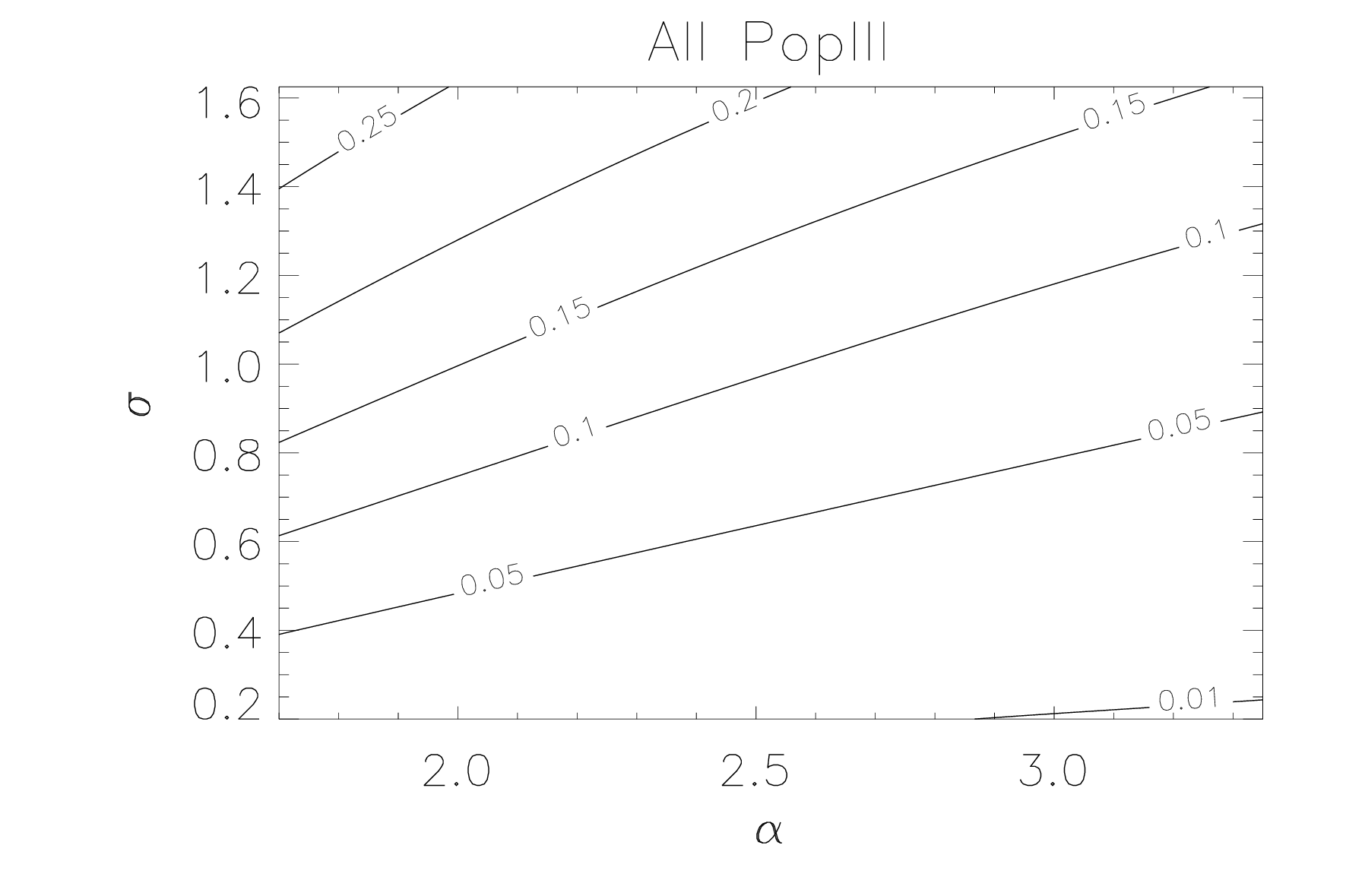}
\includegraphics[scale=0.28]{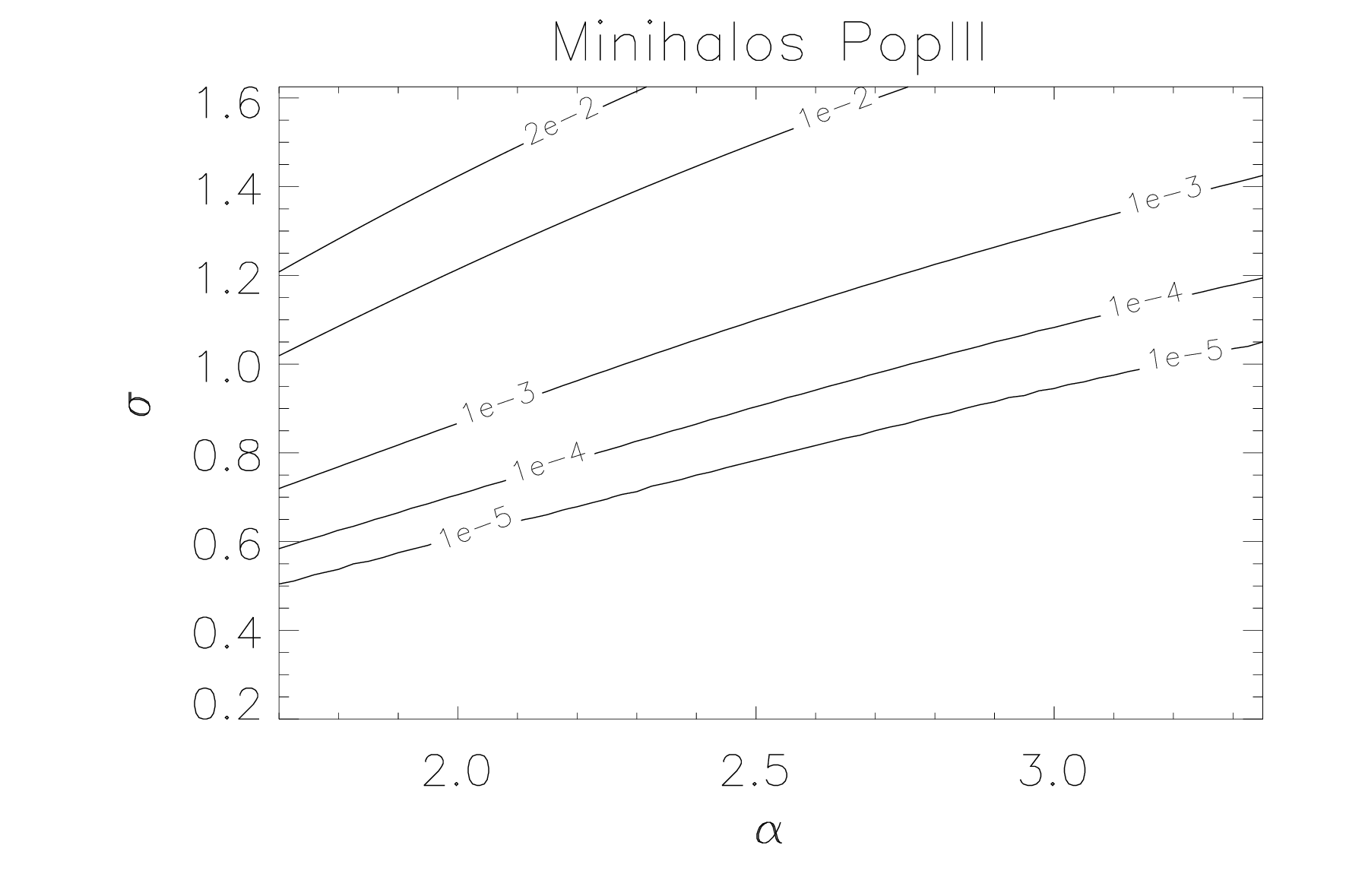}  
%%%
\caption{MW-like halo 1 ($M_1=3.15 \times 10^{12}~M_{\sun}$ at $z=0$).
  Upper left: halo projection at $z=0$ (from the low-resolution run). 
  The image size is $1.3 \times 0.97~\mathrm{Mpc^2}$.
  Upper right: formation rate of second-generation gas at
  high redshift within the comoving volume that ends up in the halo at
  $z=0$ (solid lines). This panel is the equivalent of Fig.~\ref{fig:second_gen},
  obtained for $v_{wind}=60~\mathrm{km~s^{-1}}$. For comparison,
  dotted lines show the average formation rate of second-generation
  gas from Fig.~\ref{fig:second_gen}. Smoothing over two data points
  has been applied for the solid lines. Bottom left panel: Contour
  lines for the ratio $\epsilon$ of stars enriched by Population III
  to total number of second-generation stars for halo 1. This panel is
  the equivalent of Fig.~\ref{fig:eps_std}. Bottom right panel: like
  in bottom left panel, but considering only Population III stars in
  minihalos (equivalent to Fig.~\ref{fig:eps_minihalos}).}
%%%
\label{fig:halo0} 
\end{figure}

\begin{figure} 
\includegraphics[scale=0.36]{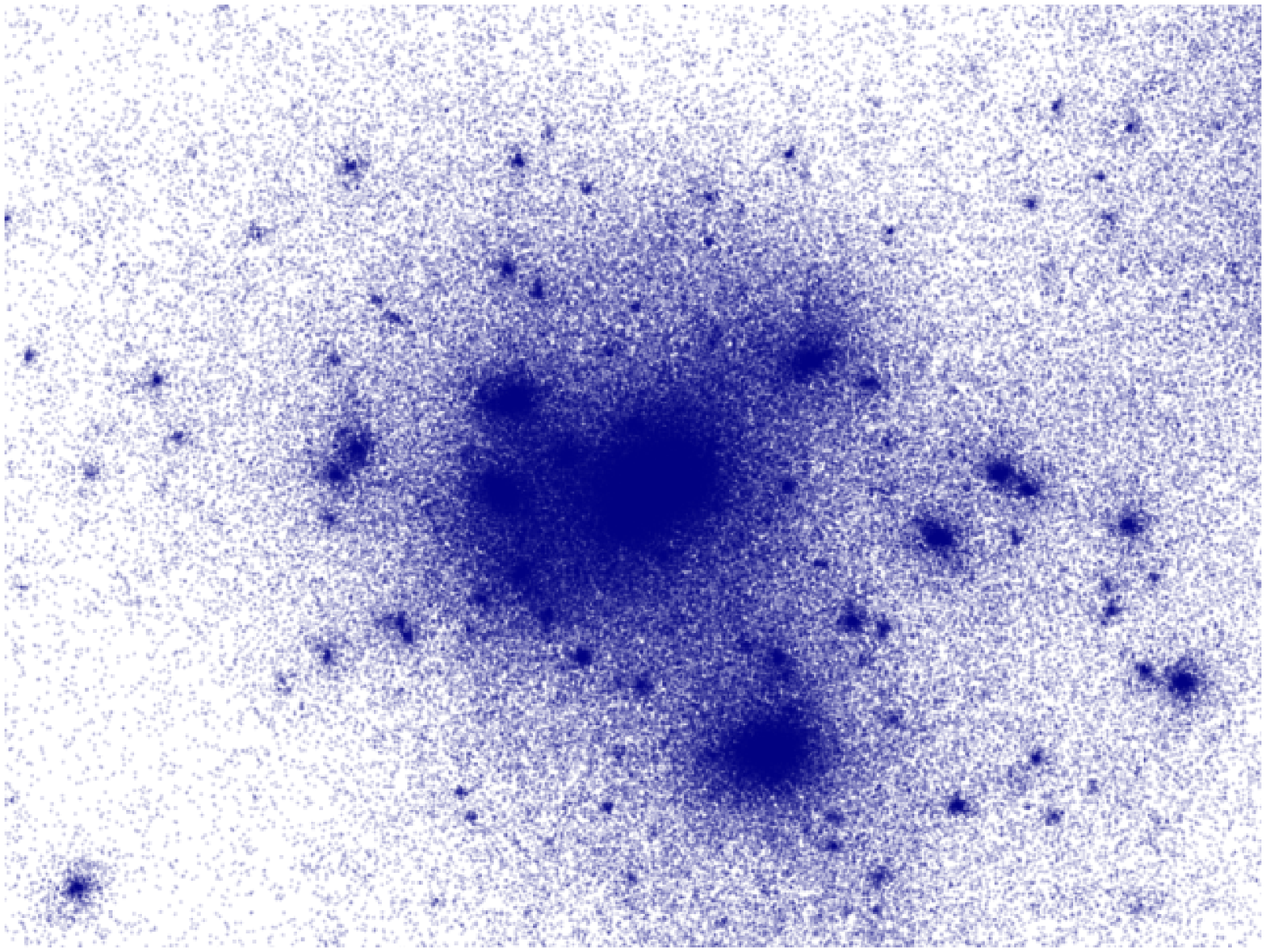}  
\includegraphics[scale=0.38]{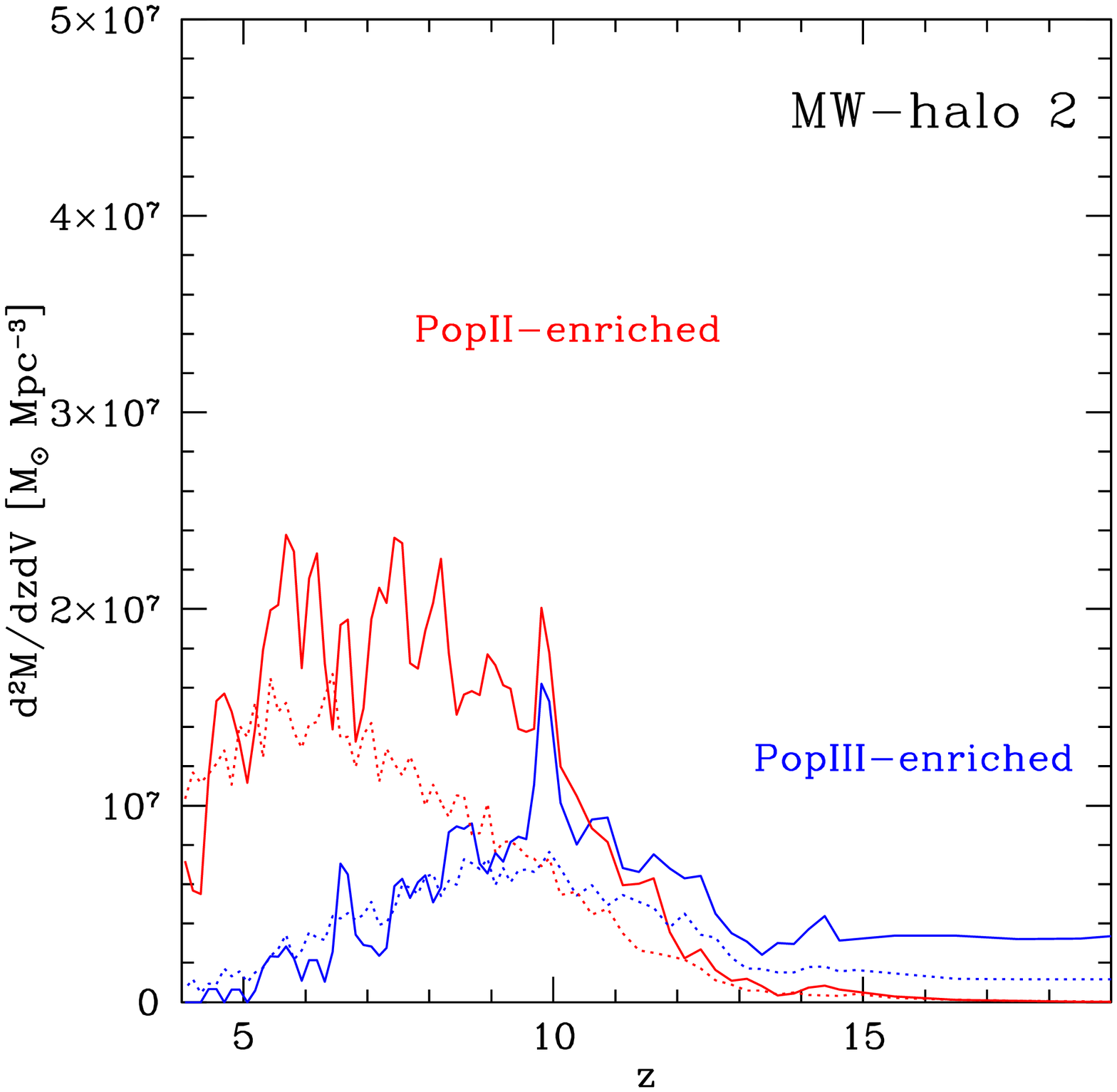}
\includegraphics[scale=0.28]{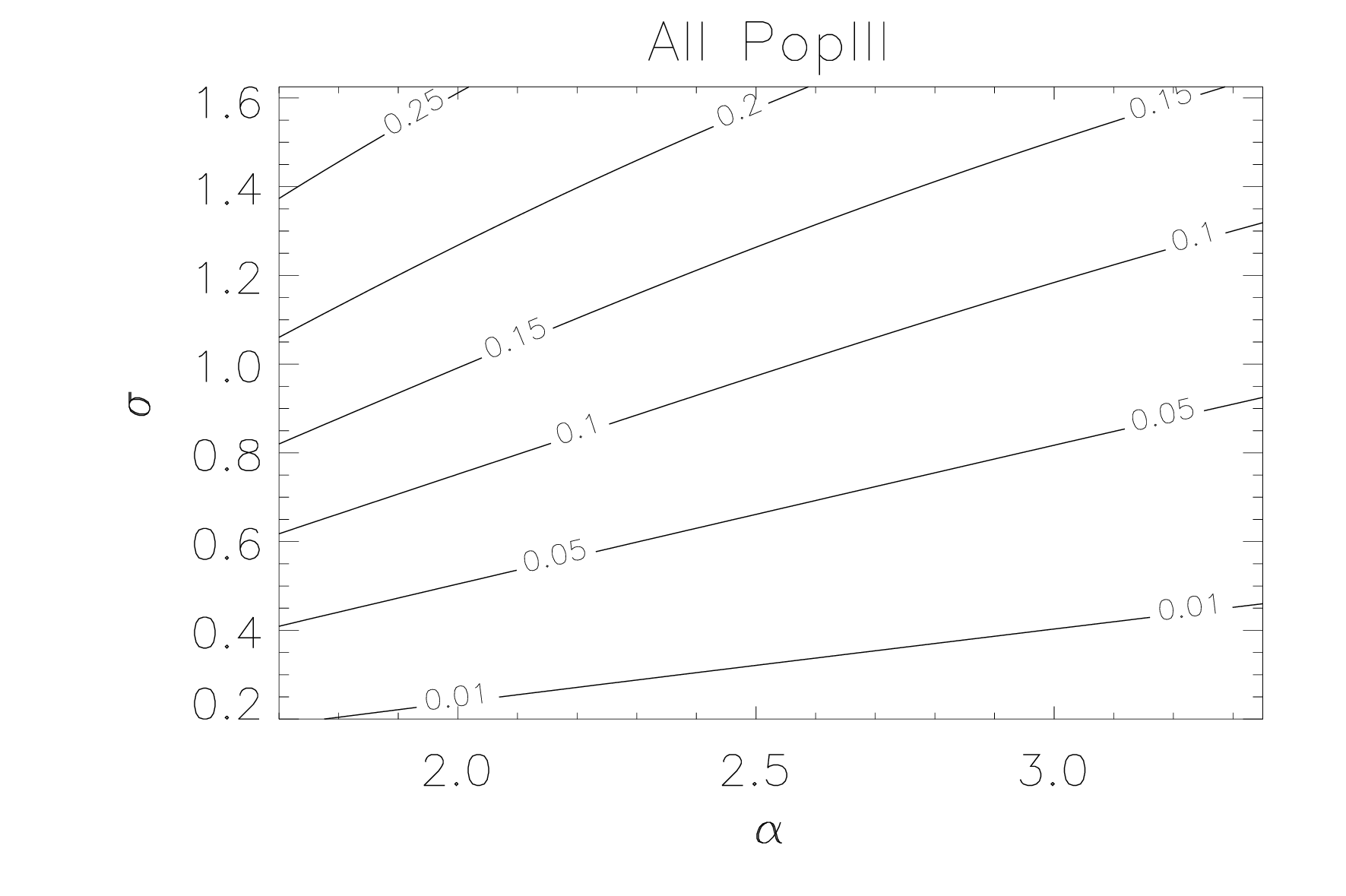}
\includegraphics[scale=0.28]{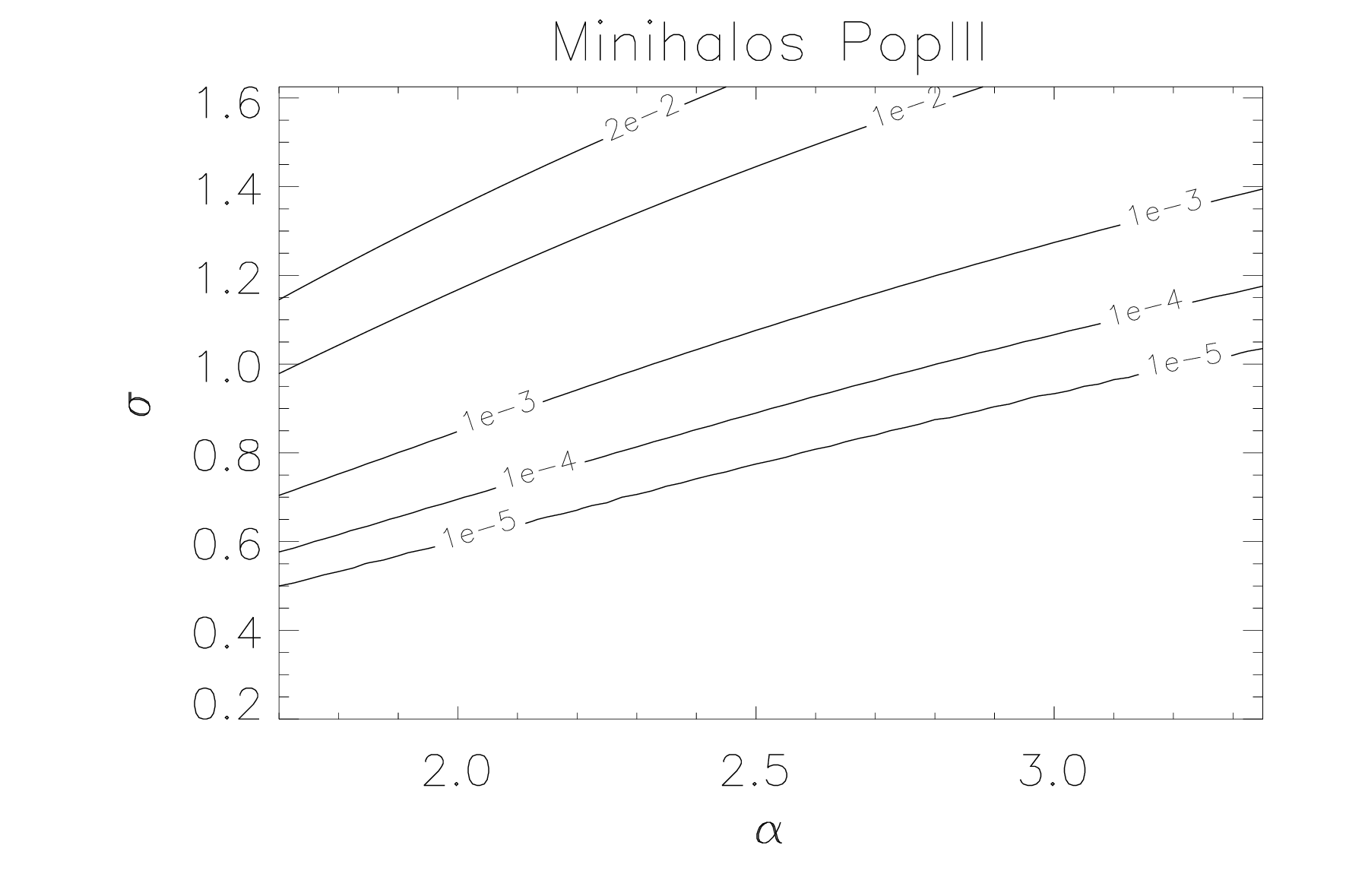}  
\caption{Same as Figure~\ref{fig:halo0} for MW-like halo 2 
($M_2=2.60 \times 10^{12}~M_{\sun}$ at $z=0$).}
\label{fig:halo1} 
\end{figure}

\begin{figure} 
\includegraphics[scale=0.36]{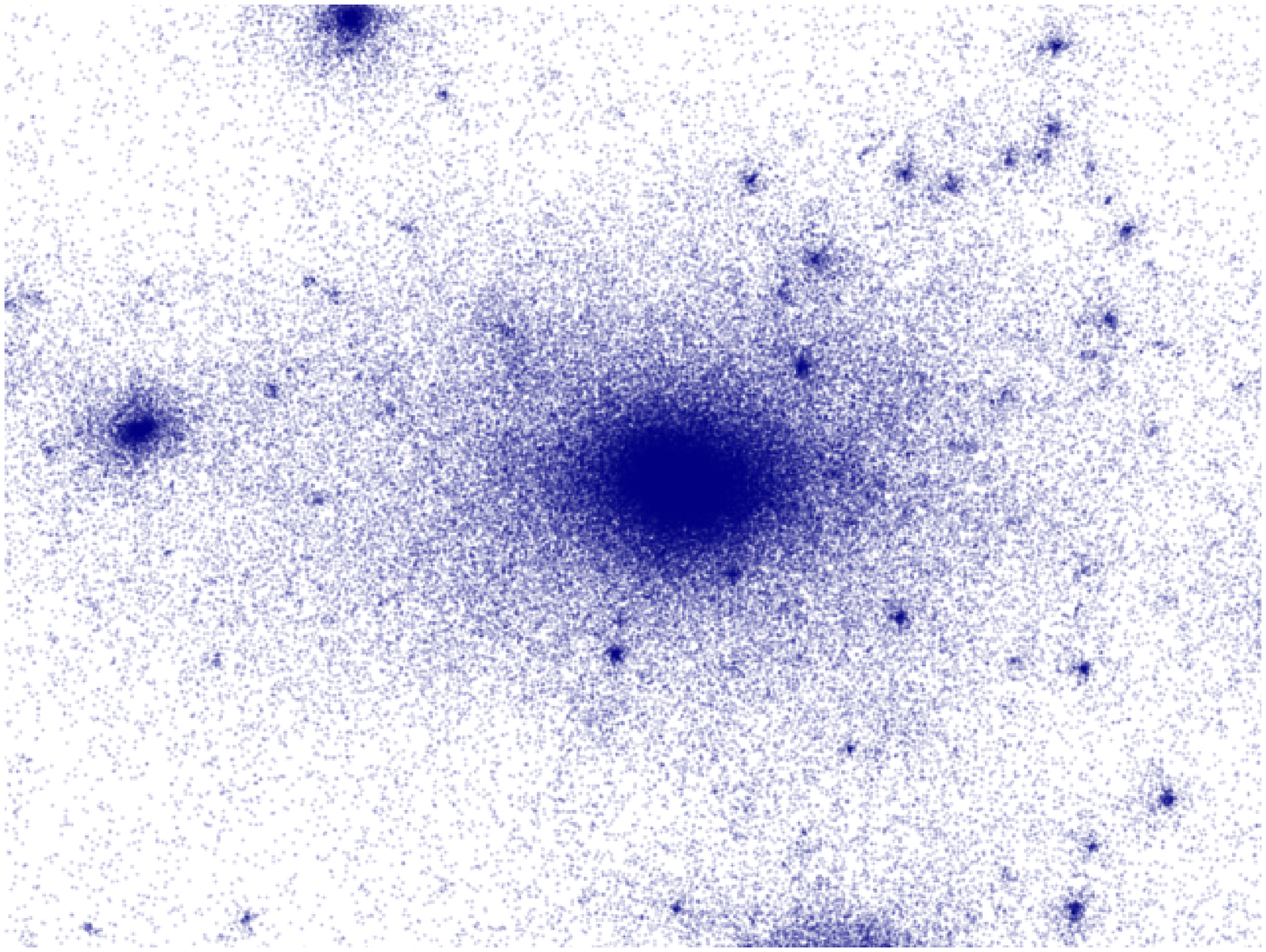}  
\includegraphics[scale=0.38]{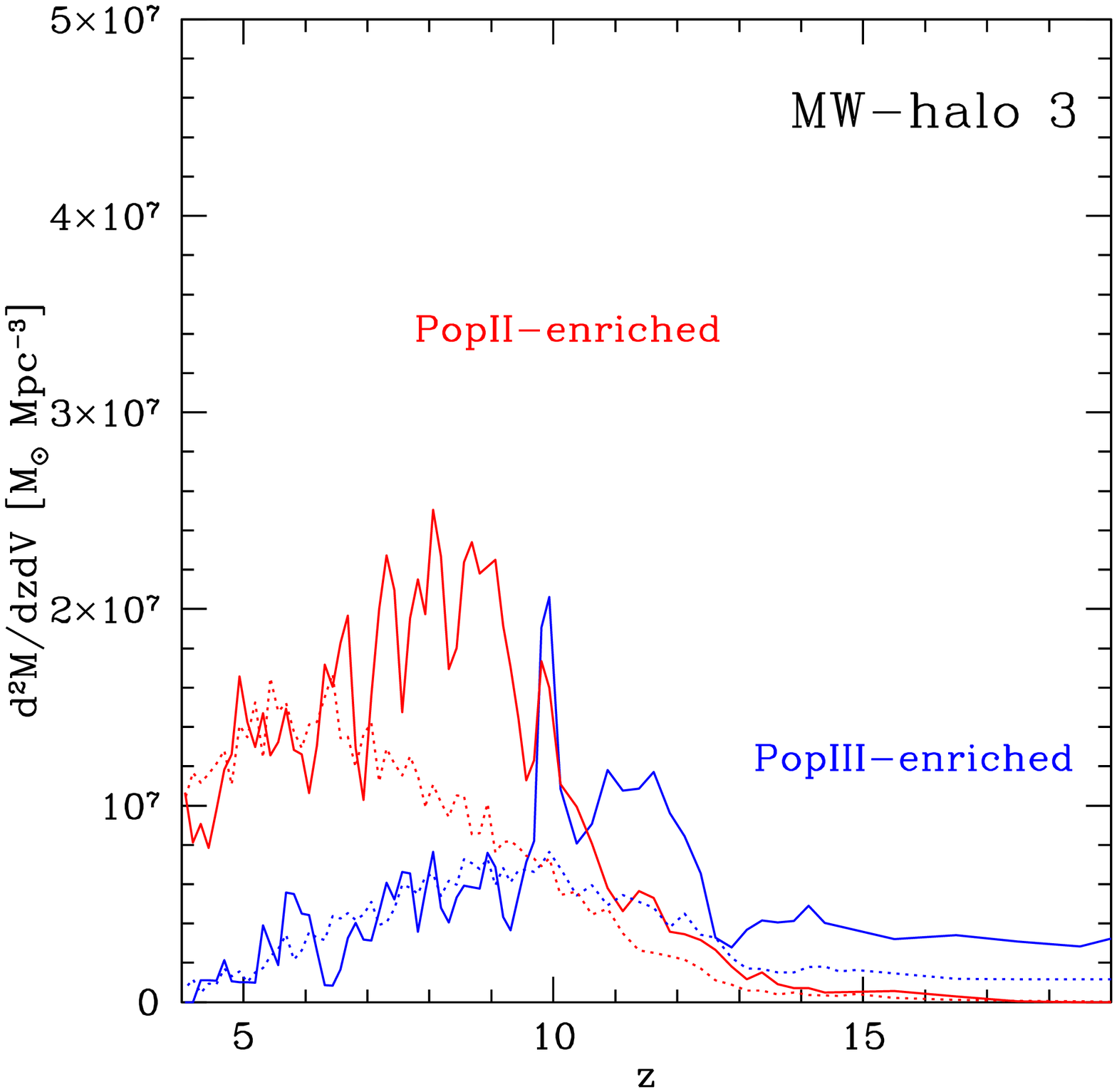}
\includegraphics[scale=0.28]{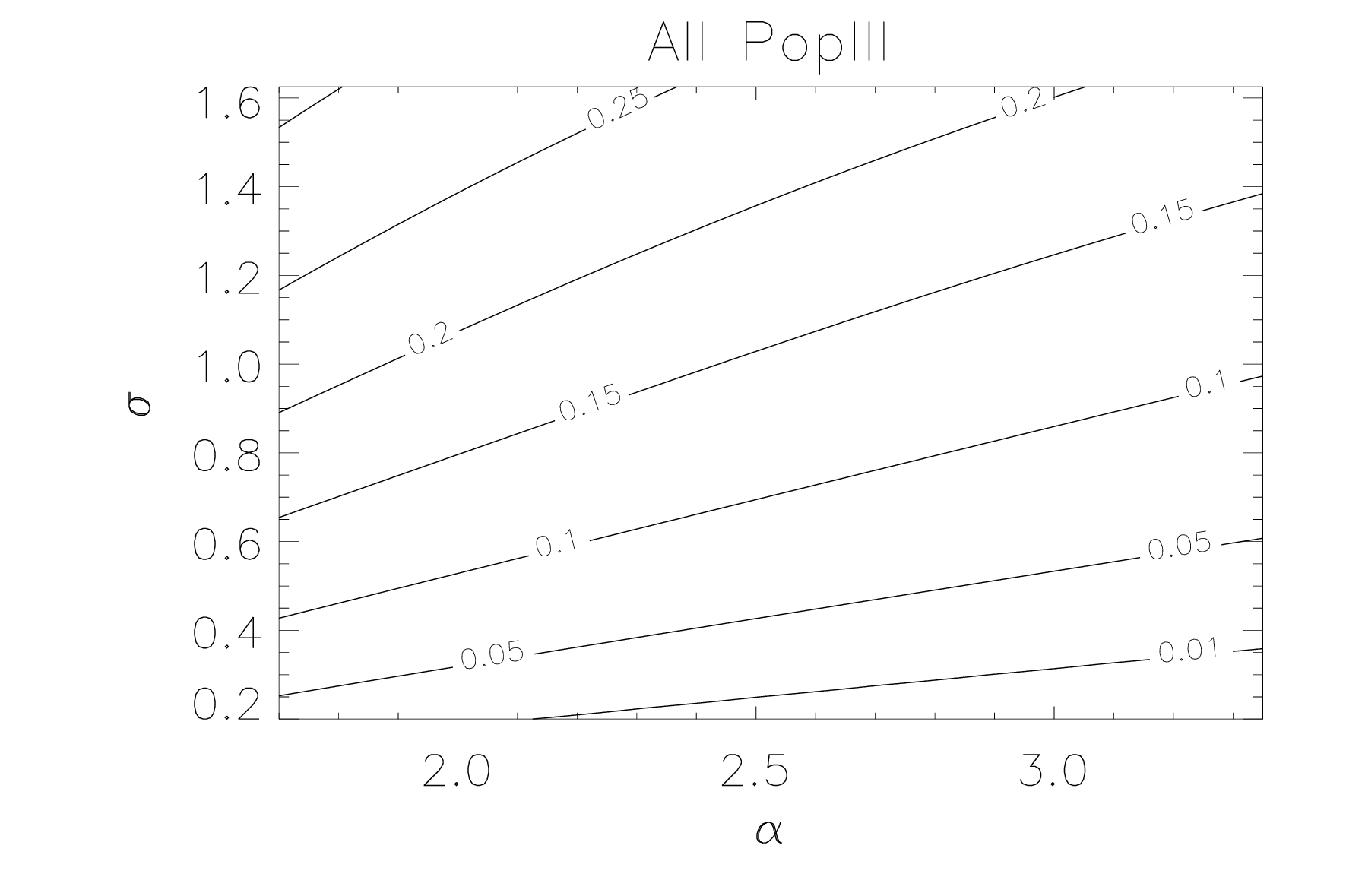}
\includegraphics[scale=0.28]{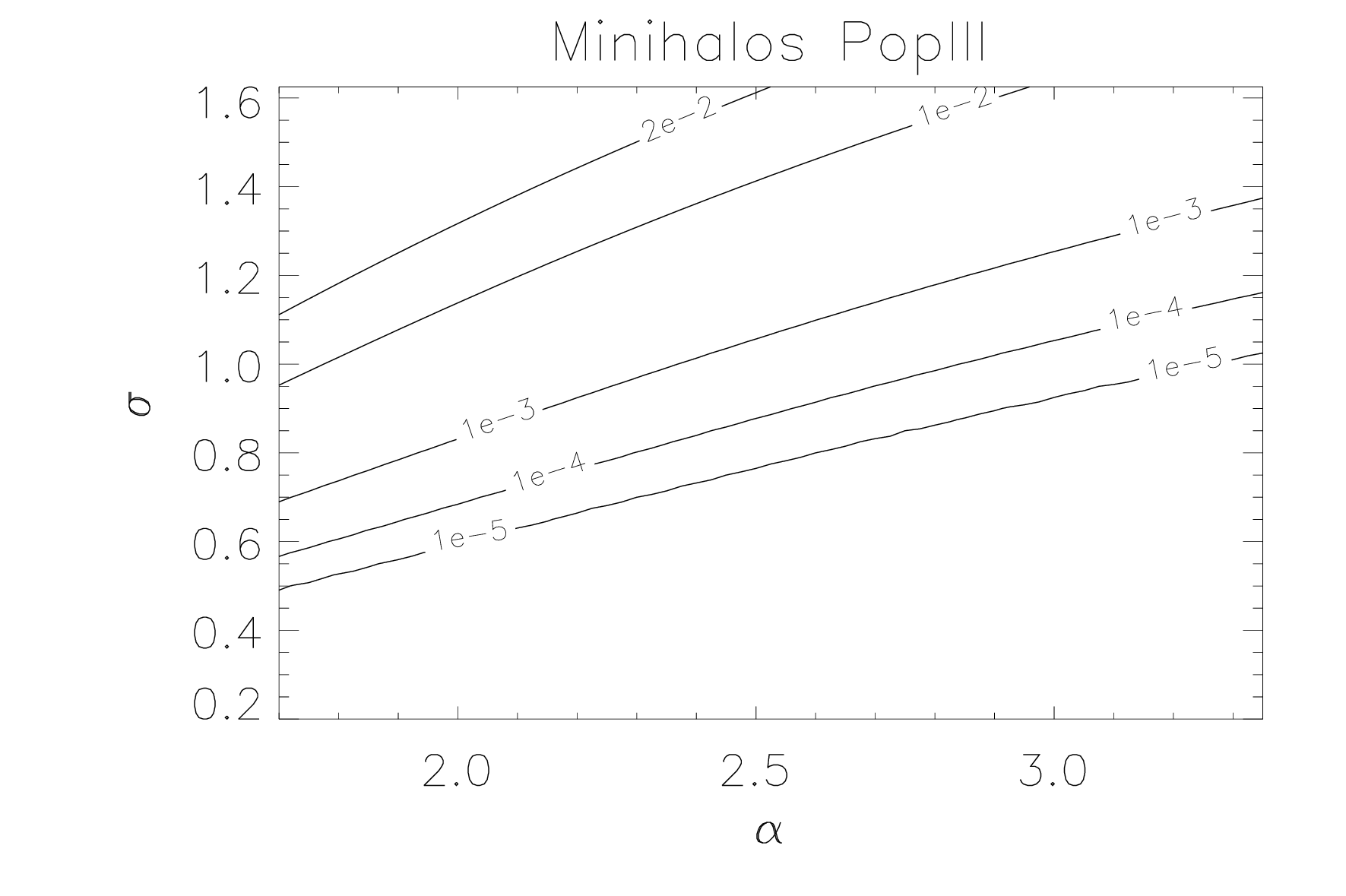}  
\caption{Same as Figure~\ref{fig:halo0} for MW-like halo 3 
($M_2=1.59 \times 10^{12}~M_{\sun}$ at $z=0$).}
\label{fig:halo2} 
\end{figure}

\begin{figure} 
\includegraphics[scale=0.38]{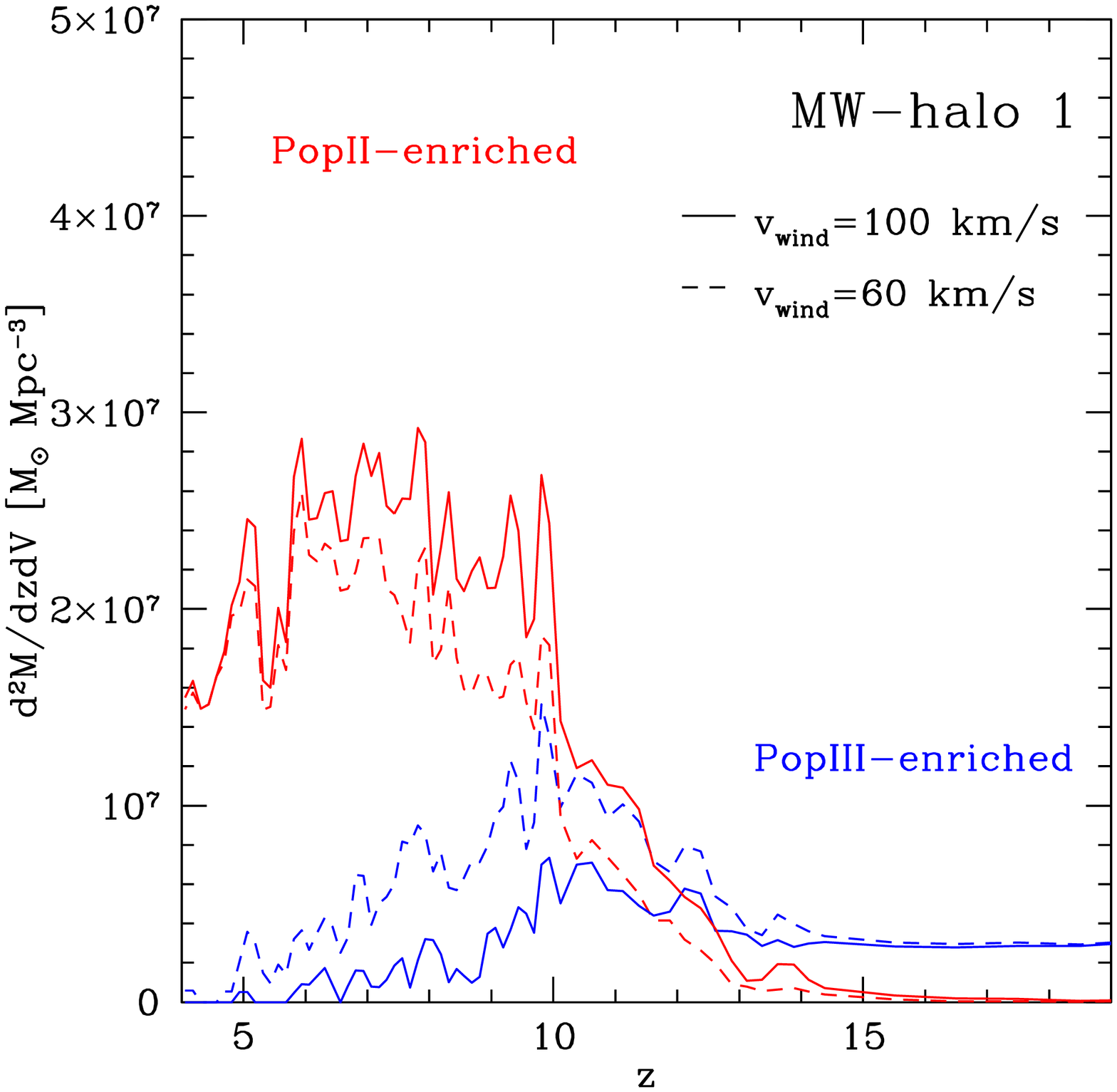}  
\includegraphics[scale=0.38]{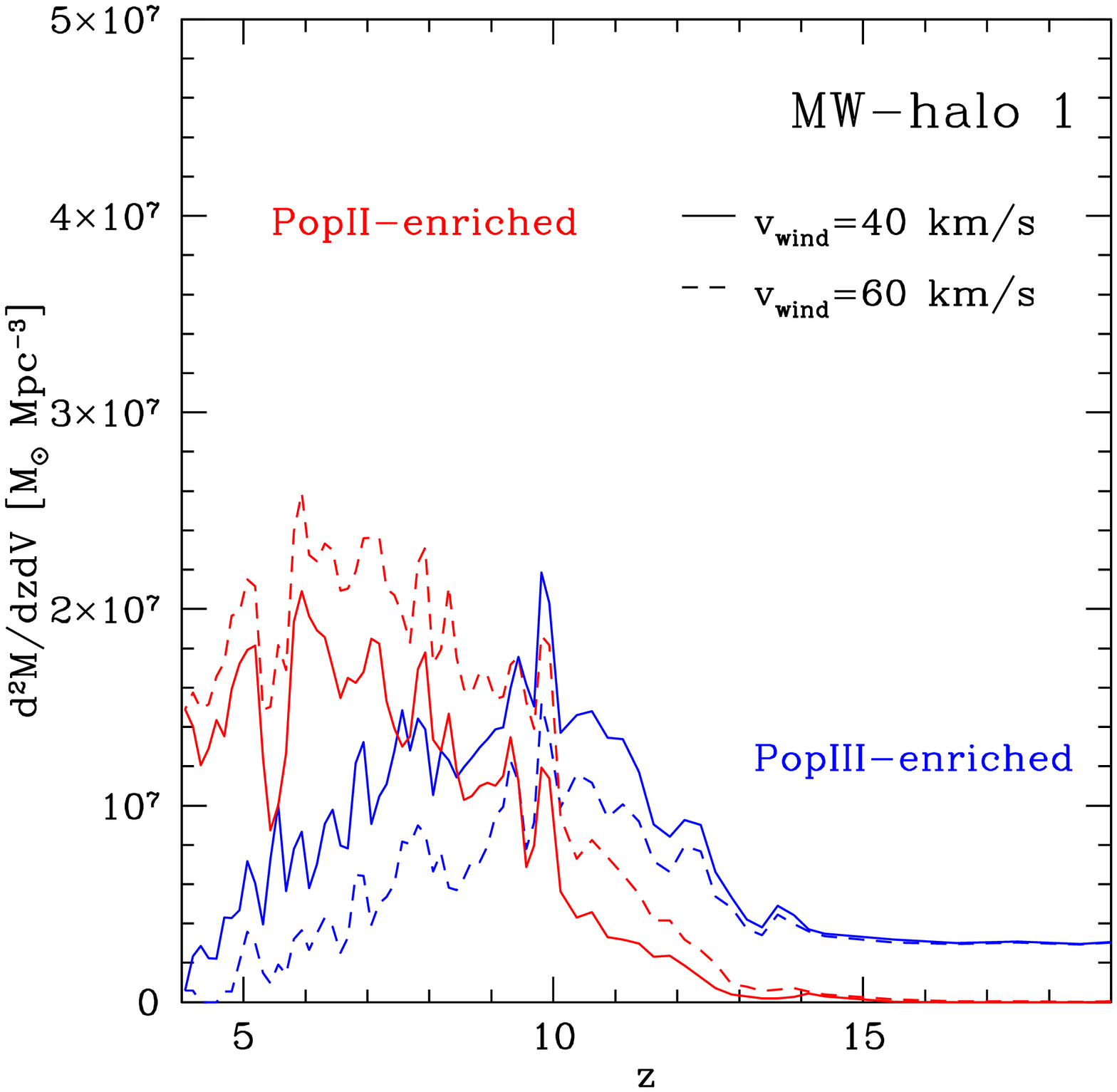}
\includegraphics[scale=0.28]{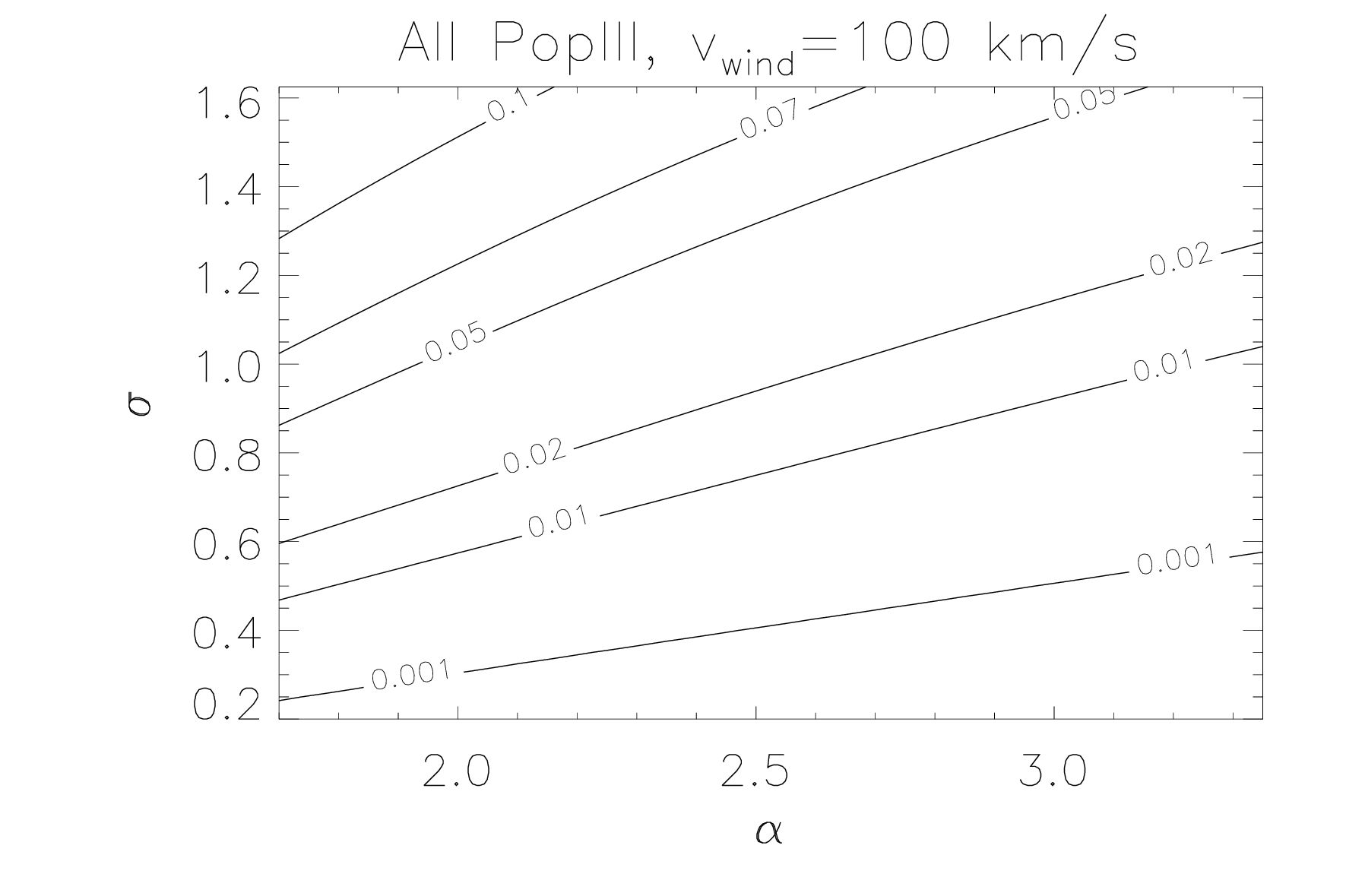}
\includegraphics[scale=0.28]{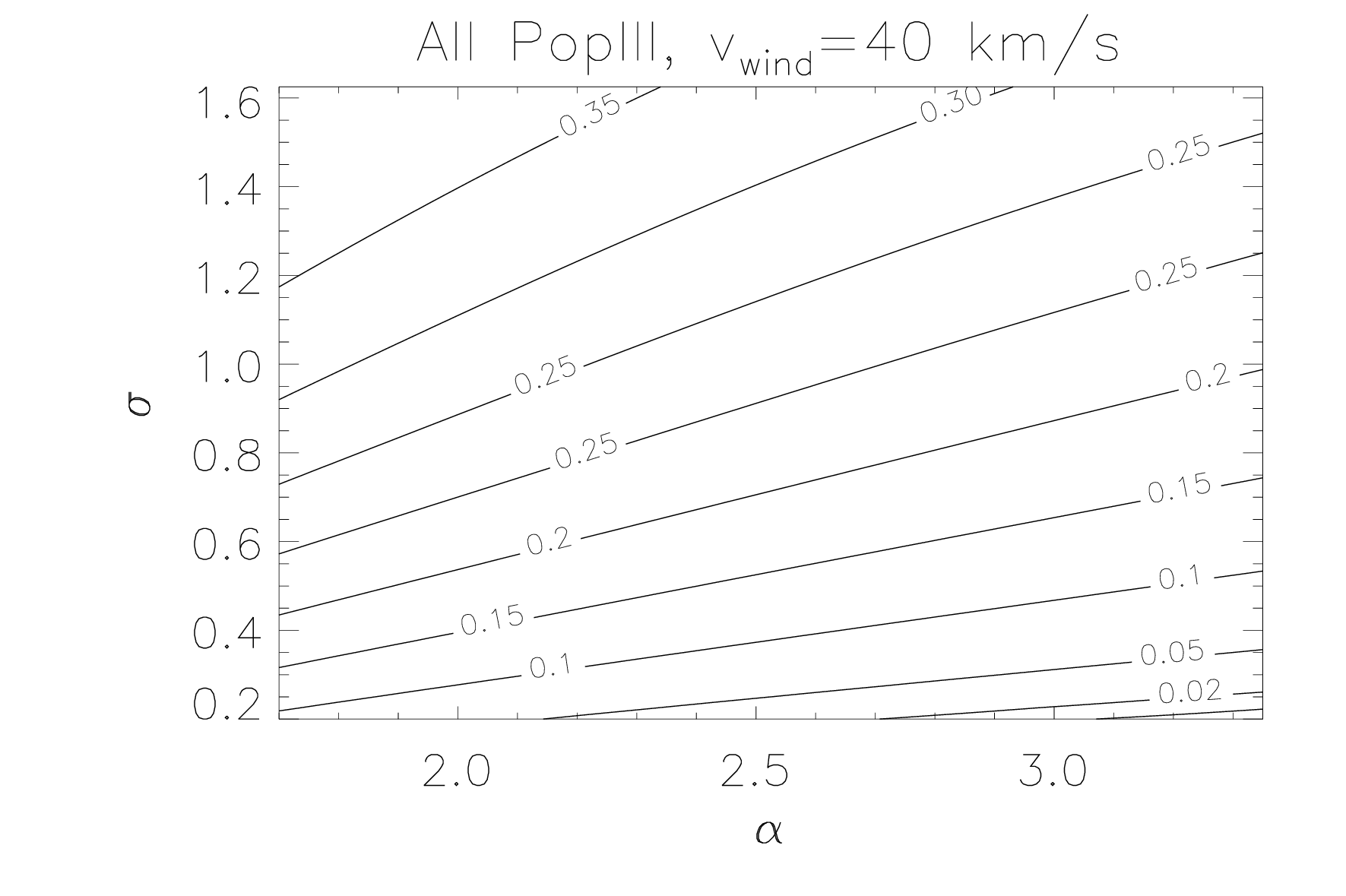}  
\caption{Influence of metal outflow speed on our modeling for 
MW-like halo 1 (see Figure~\ref{fig:halo0}). Left column shows 
results with $v_{wind}=100~\mathrm{km~s^{-1}}$, and right column 
shows $v_{wind}=40~\mathrm{km~s^{-1}}$.  Upper panels (formation i
rate of second-generation gas) show for comparison our standard model, 
$v_{wind}=60~\mathrm{km~s^{-1}}$, as dashed lines. Bottom panels show 
contour lines for the ratio $\epsilon$ of stars enriched by 
Population~III to total number of second-generation stars.}
\label{fig:halo0_wind} 
\end{figure}

From our Figures~\ref{fig:halo0}-\ref{fig:halo2}, illustrating the 
chemical enrichment of MW-like halos, we see that the comoving star
formation rate per unit redshift averages several times $10^7~
\msun$~Mpc$^{-3}$ over a redshift interval $\Delta z \approx 6$. Thus,
in a comoving volume of $20~\mathrm{Mpc^3}$, approximately equivalent
to the region that collapses into a MW-like halo, we estimate that $\sim
2.7 \times 10^9~\msun$ of second-generation gas was available to form
stars. This is $\sim 2 \%$ of the total baryonic mass of the Milky Way
today. We expect that several thousand second-generation EMP stars
could be observable today. For example, adopting a rather inefficient
star formation efficiency of 0.1\% and assuming that only
$10^{-2}$ of the second-generation stars had an initial mass below
$0.9~\msun$, we still expect a total of $2\times 10^{4}$ EMP stars
with [Fe/H] $\sim -3.5$ in the MW halo.

Finally, we have shown that the transition from Population~III to
Population~II star formation that occurred at $z\gtrsim 4$ is not too
different in the region that collapses at $z=0$ in a MW-like halo
compared to the average in the Universe
(Figures~\ref{fig:halo0}-\ref{fig:halo2}). At $z\gtrsim 10$, the
Population III star formation rate, and its associated chemical
enrichment of second generation gas are about a factor two higher
compared to the box-average. At $z=10$, metal outflows from
proto-galaxies are also more efficient, but by less than about a
factor two. However, by $z\lesssim 6$, the formation rate of
$10^{8}~M_{\sun}$ halos has returned to the average value in the
Universe. Our findings demonstrate that constrained realizations of
MW-like halos are not necessarily required to investigate the first
steps in the build-up of our own Galaxy.

\acknowledgements 

We thank Britton Smith and Massimo Stiavelli for useful comments and discussions. We
acknowledge support from the University of Colorado Astrophysical
Theory Program through grants from NASA (NNX07AG77G) and NSF
(AST07-07474).

\clearpage

%%%%%%%%%%%%%%%%%%%%%%%%%%%%%%%%% 
\bibliography{popIII}{}

\begin{thebibliography}{56}
\expandafter\ifx\csname natexlab\endcsname\relax\def\natexlab#1{#1}\fi

\bibitem[{{Abel} {et~al.}(2002){Abel}, {Bryan}, \& {Norman}}]{abel02}
{Abel}, T., {Bryan}, G.~L., \& {Norman}, M.~L. 2002, Science, 295, 93

\bibitem[{{Beers} \& {Christlieb}(2005)}]{beers05}
{Beers}, T.~C. \& {Christlieb}, N. 2005, \araa, 43, 531

\bibitem[{{Bertschinger}(2001)}]{bertschinger01}
{Bertschinger}, E. 2001, \apjs, 137, 1

\bibitem[{{Bromm} {et~al.}(2001){Bromm}, {Ferrara}, {Coppi}, \&
  {Larson}}]{bromm01a}
{Bromm}, V., {Ferrara}, A., {Coppi}, P.~S., \& {Larson}, R.~B. 2001, \mnras,
  328, 969

\bibitem[{{Bromm} \& {Larson}(2004)}]{bromm04}
{Bromm}, V. \& {Larson}, R.~B. 2004, \araa, 42, 79

\bibitem[{{Bromm} \& {Loeb}(2003)}]{brommloeb03}
{Bromm}, V. \& {Loeb}, A. 2003, \nat, 425, 812

\bibitem[{{Bromm} {et~al.}(2003){Bromm}, {Yoshida}, \& {Hernquist}}]{bromm03}
{Bromm}, V., {Yoshida}, N., \& {Hernquist}, L. 2003, \apjl, 596, L135

\bibitem[{{Christlieb} {et~al.}(2008){Christlieb}, {Sch{\"o}rck}, {Frebel},
  {Beers}, {Wisotzki}, \& {Reimers}}]{christlieb2008}
{Christlieb}, N., {Sch{\"o}rck}, T., {Frebel}, A., {Beers}, T.~C., {Wisotzki},
  L., \& {Reimers}, D. 2008, \aap, 484, 721

\bibitem[{{Clarke} \& {Bromm}(2003)}]{clarke03}
{Clarke}, C.~J. \& {Bromm}, V. 2003, \mnras, 343, 1224

\bibitem[{{Davis} {et~al.}(1985){Davis}, {Efstathiou}, {Frenk}, \&
  {White}}]{davis85}
{Davis}, M., {Efstathiou}, G., {Frenk}, C.~S., \& {White}, S.~D.~M. 1985, \apj,
  292, 371

\bibitem[{{Eisenstein} \& {Hu}(1999)}]{eisenstein99}
{Eisenstein}, D.~J. \& {Hu}, W. 1999, \apj, 511, 5

\bibitem[{{Frebel} {et~al.}(2007){Frebel}, {Johnson}, \& {Bromm}}]{frebel07}
{Frebel}, A., {Johnson}, J.~L., \& {Bromm}, V. 2007, \mnras, 380, L40

\bibitem[{{Frebel} {et~al.}(2009){Frebel}, {Johnson}, \& {Bromm}}]{frebel09}
---. 2009, \mnras, 392, L50

\bibitem[{{Freeman} \& {Bland-Hawthorn}(2002)}]{freeman02}
{Freeman}, K. \& {Bland-Hawthorn}, J. 2002, \araa, 40, 487

\bibitem[{{Furlanetto} \& {Loeb}(2003)}]{furlanetto03}
{Furlanetto}, S.~R. \& {Loeb}, A. 2003, \apj, 588, 18

\bibitem[{{Gal-Yam} {et~al.}(2009){Gal-Yam}, {Mazzali}, {Ofek}, {Nugent},
  {Kulkarni}, {Kasliwal}, {Quimby}, {Filippenko}, {Cenko}, {Chornock},
  {Waldman}, {Kasen}, {Sullivan}, {Beshore}, {Drake}, {Thomas}, {Bloom},
  {Poznanski}, {Miller}, {Foley}, {Silverman}, {Arcavi}, {Ellis}, \&
  {Deng}}]{galyam09}
{Gal-Yam}, A., {Mazzali}, P., {Ofek}, E.~O., {Nugent}, P.~E., {Kulkarni},
  S.~R., {Kasliwal}, M.~M., {Quimby}, R.~M., {Filippenko}, A.~V., {Cenko},
  S.~B., {Chornock}, R., {Waldman}, R., {Kasen}, D., {Sullivan}, M., {Beshore},
  E.~C., {Drake}, A.~J., {Thomas}, R.~C., {Bloom}, J.~S., {Poznanski}, D.,
  {Miller}, A.~A., {Foley}, R.~J., {Silverman}, J.~M., {Arcavi}, I., {Ellis},
  R.~S., \& {Deng}, J. 2009, \nat, 462, 624

\bibitem[{{Galli} \& {Palla}(1998)}]{galli98}
{Galli}, D. \& {Palla}, F. 1998, \aap, 335, 403

\bibitem[{{Greif} \& {Bromm}(2006)}]{greif06}
{Greif}, T.~H. \& {Bromm}, V. 2006, \mnras, 373, 128

\bibitem[{{Haiman} {et~al.}(1997){Haiman}, {Rees}, \& {Loeb}}]{haiman97}
{Haiman}, Z., {Rees}, M.~J., \& {Loeb}, A. 1997, \apj, 476, 458

\bibitem[{{Heger} {et~al.}(2003){Heger}, {Fryer}, {Woosley}, {Langer}, \&
  {Hartmann}}]{heger03}
{Heger}, A., {Fryer}, C.~L., {Woosley}, S.~E., {Langer}, N., \& {Hartmann},
  D.~H. 2003, \apj, 591, 288

\bibitem[{{Heger} \& {Woosley}(2002)}]{heger02}
{Heger}, A. \& {Woosley}, S.~E. 2002, \apj, 567, 532

\bibitem[{{Hoffman} \& {Ribak}(1991)}]{hoffman91}
{Hoffman}, Y. \& {Ribak}, E. 1991, \apjl, 380, L5

\bibitem[{{Ivezi{\'c}} {et~al.}(2008){Ivezi{\'c}}, {Sesar}, \&
  {Juri{\'c}}}]{ivezic_SEGUE_2008}
{Ivezi{\'c}}, {\v Z}. et~al 2008, \apj, 684, 287

\bibitem[{{Karlsson} {et~al.}(2008){Karlsson}, {Johnson}, \&
  {Bromm}}]{karlsson08}
{Karlsson}, T., {Johnson}, J.~L., \& {Bromm}, V. 2008, \apj, 679, 6

\bibitem[{{Komatsu} {et~al.}(2009){Komatsu}, {Dunkley}, {Nolta}, {Bennett},
  {Gold}, {Hinshaw}, {Jarosik}, {Larson}, {Limon}, {Page}, {Spergel},
  {Halpern}, {Hill}, {Kogut}, {Meyer}, {Tucker}, {Weiland}, {Wollack}, \&
  {Wright}}]{komatsu08}
{Komatsu}, E., {Dunkley}, J., {Nolta}, M.~R., {Bennett}, C.~L., {Gold}, B.,
  {Hinshaw}, G., {Jarosik}, N., {Larson}, D., {Limon}, M., {Page}, L.,
  {Spergel}, D.~N., {Halpern}, M., {Hill}, R.~S., {Kogut}, A., {Meyer}, S.~S.,
  {Tucker}, G.~S., {Weiland}, J.~L., {Wollack}, E., \& {Wright}, E.~L. 2009,
  \apjs, 180, 330

\bibitem[{{Komiya} {et~al.}(2009){Komiya}, {Habe}, {Suda}, \&
  {Fujimoto}}]{komiya09}
{Komiya}, Y., {Habe}, A., {Suda}, T., \& {Fujimoto}, M.~Y. 2009, \apjl, 696,
  L79

\bibitem[{{Komiya} {et~al.}(2007){Komiya}, {Suda}, {Minaguchi}, {Shigeyama},
  {Aoki}, \& {Fujimoto}}]{komiya07}
{Komiya}, Y., {Suda}, T., {Minaguchi}, H., {Shigeyama}, T., {Aoki}, W., \&
  {Fujimoto}, M.~Y. 2007, \apj, 658, 367

\bibitem[{{Kroupa}(2002)}]{kroupa2002}
{Kroupa}, P. 2002, Science, 295, 82

\bibitem[{{Larson}(2005)}]{larson05}
{Larson}, R.~B. 2005, \mnras, 359, 211

\bibitem[{{Lepp} \& {Shull}(1983)}]{lepp83}
{Lepp}, S. \& {Shull}, J.~M. 1983, \apj, 270, 578

\bibitem[{{Lynden-Bell}(1967)}]{lyndenbell67}
{Lynden-Bell}, D. 1967, \mnras, 136, 101

\bibitem[{{Machacek} {et~al.}(2003){Machacek}, {Bryan}, \& {Abel}}]{machacek03}
{Machacek}, M.~E., {Bryan}, G.~L., \& {Abel}, T. 2003, \mnras, 338, 273

\bibitem[{{Madau} {et~al.}(2001){Madau}, {Ferrara}, \& {Rees}}]{madau01}
{Madau}, P., {Ferrara}, A., \& {Rees}, M.~J. 2001, \apj, 555, 92

\bibitem[{{Naoz} {et~al.}(2006){Naoz}, {Noter}, \& {Barkana}}]{naoz06}
{Naoz}, S., {Noter}, S., \& {Barkana}, R. 2006, \mnras, 373, L98

\bibitem[{{Omukai} {et~al.}(2005){Omukai}, {Tsuribe}, {Schneider}, \&
  {Ferrara}}]{omukai05}
{Omukai}, K., {Tsuribe}, T., {Schneider}, R., \& {Ferrara}, A. 2005, \apj, 626,
  627

\bibitem[{{O'Shea} \& {Norman}(2007)}]{oshea07}
{O'Shea}, B.~W. \& {Norman}, M.~L. 2007, \apj, 654, 66

\bibitem[{{O'Shea} \& {Norman}(2008)}]{oshea08}
---. 2008, \apj, 673, 14

\bibitem[Schaerer(2003)]{schaerer03} Schaerer, D. 2003, \aap, 397, 527

\bibitem[{{Salvadori} {et~al.}(2007){Salvadori}, {Schneider}, \&
  {Ferrara}}]{salvadori07}
{Salvadori}, S., {Schneider}, R., \& {Ferrara}, A. 2007, \mnras, 381, 647

\bibitem[{{Santoro} \& {Shull}(2006)}]{santoro06}
{Santoro}, F. \& {Shull}, J.~M. 2006, \apj, 643, 26

\bibitem[{{Schneider} {et~al.}(2006){Schneider}, {Omukai}, {Inoue}, \&
  {Ferrara}}]{schneider06}
{Schneider}, R., {Omukai}, K., {Inoue}, A.~K., \& {Ferrara}, A. 2006, \mnras,
  369, 1437

\bibitem[{{Sch{\"o}rck} {et~al.}(2009){Sch{\"o}rck}, {Christlieb}, {Cohen},
  {Beers}, {Shectman}, {Thompson}, {McWilliam}, {Bessell}, {Norris},
  {Mel{\'e}ndez}, {Ram{\'{\i}}rez}, {Haynes}, {Cass}, {Hartley}, {Russell},
  {Watson}, {Zickgraf}, {Behnke}, {Fechner}, {Fuhrmeister}, {Barklem},
  {Edvardsson}, {Frebel}, {Wisotzki}, \& {Reimers}}]{schoerck2008}
{Sch{\"o}rck}, T., {Christlieb}, N., {Cohen}, J.~G., {Beers}, T.~C.,
  {Shectman}, S., {Thompson}, I., {McWilliam}, A., {Bessell}, M.~S., {Norris},
  J.~E., {Mel{\'e}ndez}, J., {Ram{\'{\i}}rez}, S., {Haynes}, D., {Cass}, P.,
  {Hartley}, M., {Russell}, K., {Watson}, F., {Zickgraf}, F., {Behnke}, B.,
  {Fechner}, C., {Fuhrmeister}, B., {Barklem}, P.~S., {Edvardsson}, B.,
  {Frebel}, A., {Wisotzki}, L., \& {Reimers}, D. 2009, \aap, 507, 817

\bibitem[{{Sheth} \& {Tormen}(1999)}]{st99}
{Sheth}, R.~K. \& {Tormen}, G. 1999, \mnras, 308, 119

\bibitem[{{Smith} {et~al.}(2009){Smith}, {Turk}, {Sigurdsson}, {O'Shea}, \&
  {Norman}}]{smith_b09}
{Smith}, B.~D., {Turk}, M.~J., {Sigurdsson}, S., {O'Shea}, B.~W., \& {Norman},
  M.~L. 2009, \apj, 691, 441

\bibitem[{{Tan} \& {McKee}(2004)}]{tan04}
{Tan}, J.~C. \& {McKee}, C.~F. 2004, \apj, 603, 383

\bibitem[{{Tegmark} {et~al.}(1997){Tegmark}, {Silk}, {Rees}, {Blanchard},
  {Abel}, \& {Palla}}]{tegmark97}
{Tegmark}, M., {Silk}, J., {Rees}, M.~J., {Blanchard}, A., {Abel}, T., \&
  {Palla}, F. 1997, \apj, 474, 1

\bibitem[{{Trenti} {et~al.}(2008){Trenti}, {Santos}, \& {Stiavelli}}]{tss08}
{Trenti}, M., {Santos}, M.~R., \& {Stiavelli}, M. 2008, \apj, 687, 1

\bibitem[{{Trenti} \& {Stiavelli}(2007)}]{ts07a}
{Trenti}, M. \& {Stiavelli}, M. 2007, \apj, 667, 38

\bibitem[{{Trenti} \& {Stiavelli}(2009)}]{ts09}
---. 2009, \apj, 694, 879

\bibitem[{{Trenti} {et~al.}(2009){Trenti}, {Stiavelli}, \& {Shull}}]{trenti09b}
{Trenti}, M., {Stiavelli}, M., \& {Shull}, J.~M. 2009, \apj, 700, 1672

\bibitem[{{Tumlinson}(2006)}]{tumlinson06}
{Tumlinson}, J. 2006, \apj, 641, 1

\bibitem[{{Tumlinson}(2007)}]{tumlinson07}
---. 2007, \apjl, 664, L63

\bibitem[{{Tumlinson}(2009)}]{tumlinson09}
---. 2009, ArXiv e-prints

\bibitem[{{Tumlinson} {et~al.}(2004){Tumlinson}, {Venkatesan}, \&
  {Shull}}]{tumlinson04}
{Tumlinson}, J., {Venkatesan}, A., \& {Shull}, J.~M. 2004, \apj, 612, 602

\bibitem[{{Weinmann} \& {Lilly}(2005)}]{weinmann05}
{Weinmann}, S.~M. \& {Lilly}, S.~J. 2005, \apj, 624, 526

\bibitem[{{Yoshida} {et~al.}(2003){Yoshida}, {Abel}, {Hernquist}, \&
  {Sugiyama}}]{yoshida03}
{Yoshida}, N., {Abel}, T., {Hernquist}, L., \& {Sugiyama}, N. 2003, \apj, 592,
  645

\bibitem[{{Yoshida} {et~al.}(2006){Yoshida}, {Omukai}, {Hernquist}, \&
  {Abel}}]{yoshida06}
{Yoshida}, N., {Omukai}, K., {Hernquist}, L., \& {Abel}, T. 2006, \apj, 652, 6

\end{thebibliography}

\clearpage

%%%%%%%%%%%%%%%%%%%%%%%%%%%%%%%%%%%%%%%%%%%%%
%%%%%%%%%%%%%%%%%%%%%%%%%%%%%%%%%%%%%%%%%%%

%%%%%%%%%%%%%%%%%%%%%%%%%%%%%%%%%%%%%%%%%%%%%
%%%%%%%%%%%%%%%%%%%%%%%%%%%%%%%%%%%%%%%%%%%

%%%%%%%%%%%%%%%%%%%%%%%%%%%%%%%%%%%%%%%%%%%%%
%%%%%%%%%%%%%%%%%%%%%%%%%%%%%%%%%%%%%%%%%%%

%%%%%%%%%%%%%%%%%%%%%%%%%%%%%%%%%%%%%%%%%%%%%
%%%%%%%%%%%%%%%%%%%%%%%%%%%%%%%%%%%%%%%%%%%

%%%%%%%%%%%%%%%%%%%%%%%%%%%%%%%%%%%%%%%%%%%%%
%%%%%%%%%%%%%%%%%%%%%%%%%%%%%%%%%%%%%%%%%%%

%%%%%%%%%%%%%%%%%%%%%%%%%%%%%%%%%%%%%%%%%%%%%
%%%%%%%%%%%%%%%%%%%%%%%%%%%%%%%%%%%%%%%%%%%

%%%%%%%%%%%%%%%%%%%%%%%%%%%%%%%%%%%%%%%%%%%%%
%%%%%%%%%%%%%%%%%%%%%%%%%%%%%%%%%%%%%%%%%%%

%%%%%%%%%%%%%%%%%%%%%%%%%%%%%%%%%%%%%%%%%%%%%%%%%%
\begin{table} \begin{center} \caption{Required sample size of EMP stars with $Z \leqslant 10^{-3.4}~ Z_{\sun}$\label{tab:sample_size}}
%%%%%%%%%%%%%%%%%%%
\smallskip
\smallskip
\smallskip
\begin{tabular}{cccc}
\tableline\tableline
\smallskip $\alpha$ & $\sigma$ & $N_{min-AllPopIII}$ & $N_{min-MinihaloPopIII}$ \\
\tableline
\smallskip
1.7 & 0.3 & 118~~~~274 & $>10^7~~~~>10^7$  \\
1.7 & 0.6 &  65~~~~95 & $1.35 \times 10^5~~~~6.63 \times 10^4$ \\
1.7 & 1.0 &  41~~~~49 & 1775~~~~1008  \\
2.5 & 0.3 &  130~~~~449 & $>10^7~~~~>10^7$  \\
2.5 & 0.6 &   99~~~~202&  $>10^7 ~~~~>10^7$  \\
2.5 & 1.0 &   62~~~~87 &  $5.09 \times 10^5~~~~2.54\times 10^4$ \\
3.35 & 0.3 & 135~~~~588 & $>10^7~~~~>10^7$ \\
3.35 & 0.6 & 120~~~~344 & $>10^7~~~~>10^7$ \\
3.35 & 1.0 & 86~~~~151 &  $4.61 \times 10^6~~~~2.30\times 10^6$ \\
\tableline
\end{tabular}
%% Any table notes must follow the \end{tabular} command.
\tablecomments{Minimum number ($N_{min}$, see Equation~\ref{eq:min_n}) of
  second-generation stars needed to
  rule out at a confidence level of 99\% the hypothesis that a fraction 50\% of
  Population~III stars exploded as PISN (third column) as a function
  of the second-generation IMF parameters $\alpha$ and $\sigma$. The
  first value of $N_{min}$ refers to the box average, the second value
  to MW-like halo 1. The
  last column reports the minimum number of second-generation stars
  needed to rule out, at the same confidence level, PISN in minihalos
  Population~III stars. For some of the IMF parameter space, the number
  of stars needed exceeds $10^7$. The values of the characteristic IMF
  width $\sigma$ reported here are those assumed in \citet{tumlinson06}.}
%%%%
\end{center}
\end{table}

\end{document}